\newif\ifarxiv %
\arxivtrue  %

\ifarxiv %
    \documentclass{article}
    \usepackage{arxiv}
    \usepackage{multirow}
    \usepackage[utf8]{inputenc} %
    \usepackage[T1]{fontenc}    %
    \usepackage{hyperref}       %
    \usepackage{url}            %
    \usepackage{booktabs}       %
    \usepackage{amsfonts}       %
    \usepackage{nicefrac}       %
    \usepackage{microtype}      %
    \usepackage{doi}
    \title{Large Language Models Meet User Interfaces: The Case of Provisioning Feedback}  

    \author{
        Stanislav Pozdniakov \\
        School of Electrical Engineering and Computer Science \\
        The University of Queensland \\
        St Lucia, QLD, 4072, Australia \\
	\texttt{s.pozdniakov@uq.edu.au} \\
    \And
        Jonathan Brazil \\
        Institute for Teaching and Learning Innovation \\
        The University of Queensland \\
        St Lucia, QLD, 4072, Australia \\
        \texttt{j.brazil@uq.edu.au} \\
    \And
        Solmaz Abdi \\
        School of Electrical Engineering and Computer Science \\
        The University of Queensland \\
        St Lucia, QLD, 4072, Australia \\
    \And
	Aneesha Bakharia \\
        School of Electrical Engineering and Computer Science \\
        The University of Queensland \\
        St Lucia, QLD, 4072, Australia \\
	\texttt{a.bakharia1@uq.edu.au} \\
    \And
	Shazia Sadiq \\
        School of Electrical Engineering and Computer Science \\
        The University of Queensland \\
        St Lucia, QLD, 4072, Australia \\
	\texttt{shazia@eecs.uq.edu.au } \\
    \And
	Dragan Ga{\v s}evi{\'c} \\
        Centre for Learning Analytics \\
        Faculty of Information Technology \\
        Monash University \\
        Melbourne, VIC, 3800, Australia \\
        \texttt{dragan.gasevic@monash.edu} \\
    \And
	Paul Denny \\
        School of Computer Science \\
        University of Auckland  \\
        Auckland, New Zealand \\
        \texttt{paul@cs.auckland.ac.nz} \\
    \And
        Hassan Khosravi \\
        School of Electrical Engineering and Computer Science \\
        The University of Queensland \\
        St Lucia, QLD, 4072, Australia \\
	\texttt{h.khosravi@uq.edu.au} \\
}
    \renewcommand{\shorttitle}{Large Language Models Meet User Interfaces}
    
\else %
    \documentclass[a4paper,fleqn]{./els-cas-template/cas-sc} %
\fi

\usepackage{graphicx} %
\usepackage[dvipsnames]{xcolor}
\usepackage{longtable}
\usepackage{tabu}
\usepackage{listings}
\usepackage[authoryear]{natbib}
\usepackage{float}

\definecolor{criterion1}{HTML}{4B7F5A}
\definecolor{criterion2}{HTML}{CEA820}
\definecolor{criterion3}{HTML}{EC1865}
\definecolor{criterion4}{HTML}{0F6CBD}

\begin{document}

\ifarxiv %
\else %
    \let\WriteBookmarks\relax
    \def\floatpagepagefraction{1}
    \def\textpagefraction{.001}

    \shorttitle{Large Language Models Meet User Interfaces} %
    \shortauthors{Anon et~al.} %
    \title[mode = title]{Large Language Models Meet User Interfaces: The Case of Provisioning Feedback}  
    
\fi

\ifarxiv %
    \maketitle

\begin{abstract}
Incorporating Generative Artificial Intelligence (GenAI), especially Large Language Models (LLMs), into educational settings presents valuable opportunities to boost the efficiency of educators and enrich the learning experiences of students. A significant portion of the current use of LLMs by educators has involved using conversational user interfaces (CUIs), such as chat windows, for functions like generating educational materials or offering feedback to learners. 
The ability to engage in real-time conversations with LLMs, which can enhance educators' domain knowledge across various subjects, has been of high value. However, it also presents challenges to LLMs' widespread, ethical, and effective adoption. 
Firstly, educators must have a degree of expertise, including tool familiarity, AI literacy and prompting to effectively use CUIs, which can be a barrier to adoption. Secondly, the open-ended design of CUIs makes them exceptionally powerful, which raises ethical concerns, particularly when used for high-stakes decisions like grading. Additionally, there are risks related to privacy and intellectual property, stemming from the potential unauthorised sharing of sensitive information. Finally, CUIs are designed for short, synchronous interactions and often struggle and hallucinate when given complex, multi-step tasks (e.g., providing individual feedback based on a rubric on a large scale).
To address these challenges, we explored the benefits of transitioning away from employing LLMs via CUIs to the creation of applications with user-friendly interfaces that leverage LLMs through API calls. We first propose a framework for pedagogically sound and ethically responsible incorporation of GenAI into educational tools, emphasizing a human-centered design. We then illustrate the application of our framework to the design and implementation of a novel tool called Feedback Copilot, which enables instructors to provide students with personalized qualitative feedback on their assignments in classes of any size. 
An evaluation involving the generation of feedback from two distinct variations of the Feedback Copilot tool, using numerically graded assignments from 338 students, demonstrates the viability and effectiveness of our approach. Our findings have significant implications for GenAI application researchers, educators seeking to leverage accessible GenAI tools, and educational technologists aiming to transcend the limitations of conversational AI interfaces, thereby charting a course for the future of GenAI in education.
\end{abstract}

\ifarxiv %

\section*{\centering{Highlights}}
\newenvironment{highlights}
    {
    \begin{itemize}
    
    }
    { 
    \end{itemize}
    }

\else %
    \begin{graphicalabstract}
    \includegraphics[width=1\textwidth]{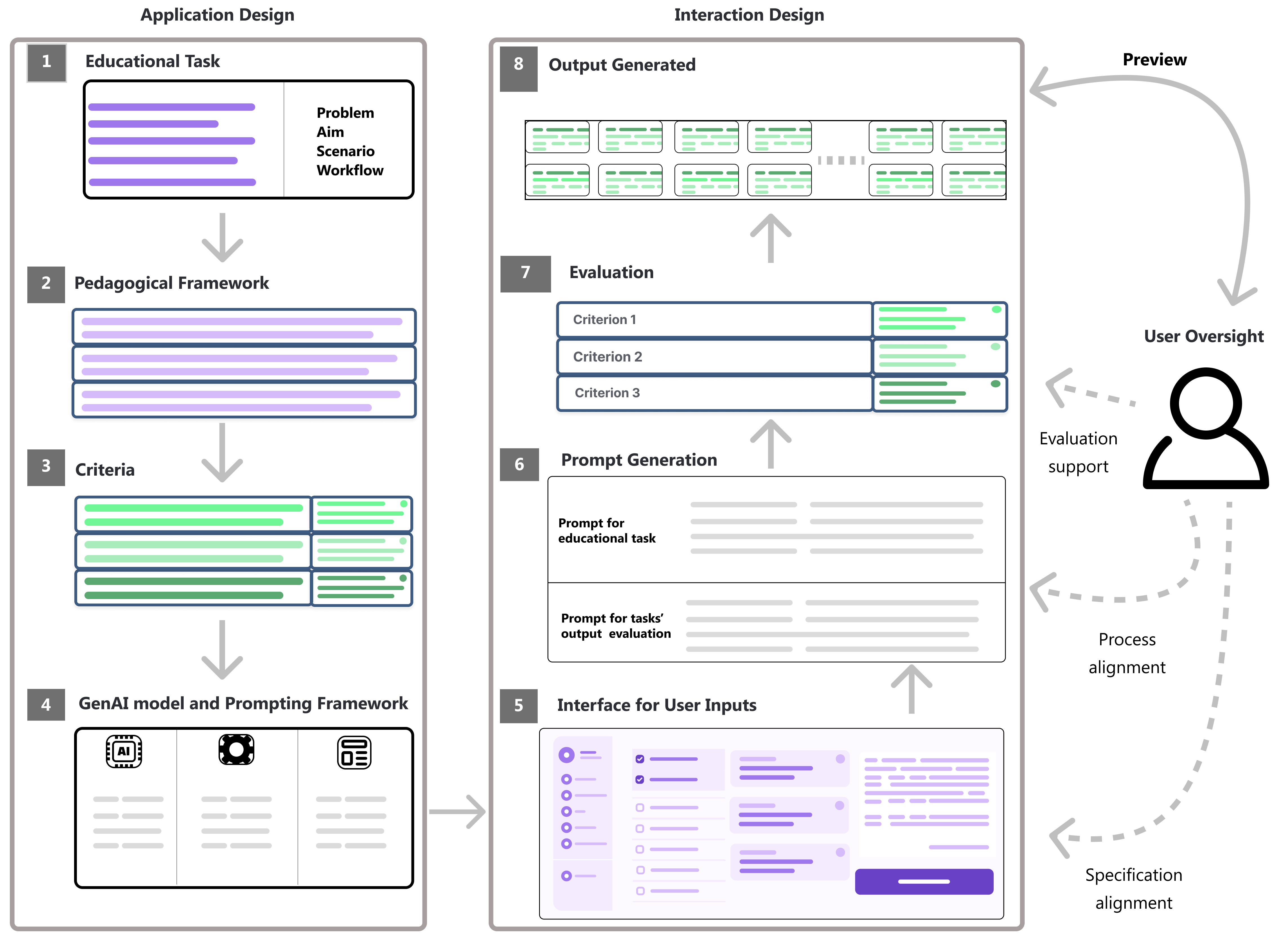}
    \end{graphicalabstract}
\fi

\begin{highlights}
\item The rise of Generative Artificial Intelligence (GenAI) particularly through Large Language Models (LLM) shows promise in enhancing educator productivity and student learning.
\item Utilizing GenAI via conversational user interfaces (CUIs) is hindered by the need for expertise in prompting, posing potential risks related to privacy and ethics, and inability to perform complex instructional activities.
\item We investigate the potential of embedding GenAI into user-centric applications to address these barriers and propose a framework that outlines a pedagogically sound and ethically responsible approach for the integration.
\item An evaluation on the assignments of 338 students demonstrates the practical application of our framework and the novel tool called `Feedback Copilot', which produces qualitative feedback and is designed to be used by instructors
\item Our research underscores the potential for GenAI-enhanced educational tools, encouraging further exploration into user-centric GenAI applications.
\end{highlights}

\else %

\ifarxiv %

\section*{\centering{Highlights}}
\newenvironment{highlights}
    {
    \begin{itemize}
    
    }
    { 
    \end{itemize}
    }

\else %
    \begin{graphicalabstract}
    \includegraphics[width=1\textwidth]{FINAL_Framework_Overall.png}
    \end{graphicalabstract}
\fi

\begin{highlights}
\item The rise of Generative Artificial Intelligence (GenAI) particularly through Large Language Models (LLM) shows promise in enhancing educator productivity and student learning.
\item Utilizing GenAI via conversational user interfaces (CUIs) is hindered by the need for expertise in prompting, posing potential risks related to privacy and ethics, and inability to perform complex instructional activities.
\item We investigate the potential of embedding GenAI into user-centric applications to address these barriers and propose a framework that outlines a pedagogically sound and ethically responsible approach for the integration.
\item An evaluation on the assignments of 338 students demonstrates the practical application of our framework and the novel tool called `Feedback Copilot', which produces qualitative feedback and is designed to be used by instructors
\item Our research underscores the potential for GenAI-enhanced educational tools, encouraging further exploration into user-centric GenAI applications.
\end{highlights}

    \begin{keywords}
        Artificial Intelligence \sep
        Large Language Models \sep
        Generative Artificial Intelligence \sep 
        Interfaces \sep
        Feedback \sep
        Learning Analytics
    \end{keywords}
    \maketitle
\fi

\section{Introduction}  

We are increasingly seeing a wide interest in the applications of Generative Artificial Intelligence (GenAI) and, particularly, Large Language Models (LLMs), to help educators and instructors improve their practices in a wide range of pedagogical scenarios. 
Examples include the development of course content \citep{dickeyModelIntegratingGenerative2023,dennyCanWeTrust2023}, e.g., creation of multiple-choice questions \citep{bulathwelaScalableEducationalQuestion2023,mooreAssessingQualityMultipleChoice2023} and AI-generated learning videos clips \citep{leikerGenerativeAILearning2023}; integration with Intelligent Tutoring Systems, e.g., automatic hint generation \citep{pardosLearningGainDifferences2023}; incorporating GenAI innovations into adaptive learning systems and for personalization, i.e., producing on-demand interactive worked examples \citep{jury2024evaluating}, and utilising LLMs to create initial templates to help students in learnersourcing \citep{khosraviLearnersourcingAgeAI2023}; applications for assessment and evaluation, i.e, utilising LLMs to improve tutors' feedback literacy \citep{linUsingLargeLanguage} or using LLMs to deliver automatic feedback for students' assignments \citep{hanFABRICAutomatedScoring2023}. 

A significant portion of the current use of GenAI has been through conversational interfaces, where users engage with LLMs via a series of synchronous sequential messages. While the capacity to hold an in-depth, real-time dialogue with LLMs is particularly compelling and provides exciting opportunities, this very capability poses obstacles to their widespread, ethical, and effective use \citep{choiAreLLMsUseful2023}. One hurdle is the \textbf{\textit{AI literacy}} barrier as leveraging these models effectively requires both students and educators to possess knowledge of their capabilities and limitations \citep{longWhatAILiteracy2020}, as well as sophisticated prompt engineering skills \citep{whitePromptPatternCatalog2023,cainPromptingChangeExploring2023,denny2023conversing}. The necessity for AI proficiency is compounded by the essential need to integrate \textit{\textbf{educational theories}} and principles within these prompts \citep{limGenerativeAIFuture2023,mollickAssigningAISeven2023}, ensuring the generated content meets best pedagogical practices which can pose a significant barrier to effective implementation.

Challenges around \textbf{\textit{user autonomy}} present another layer of complexity. Firstly, in contrast to conventional educational tools built for precise tasks that are equipped with helpful guides and offer step-by-step guidance, CUIs require users to determine their own objectives and master the intricacies involved \citep{arawjoChainForgeOpensourceVisual2023,suhSensecapeEnablingMultilevel2023,angertSpellburstNodebasedInterface2023}. 
For instance, an educator using an LLM to provide feedback on a student’s assignment might face the challenge of articulating specific, actionable advice without the structured prompts provided by traditional grading rubrics. The educator needs to independently navigate the LLM’s capabilities to effectively prompt the LLM to formulate such feedback. This shift from structured to open-ended interfaces can leave educators, who are used to more guided systems, struggling to use CUIs effectively. Moreover, GenAI CUIs often lack access to institutional educational data, such as course materials, student interaction logs, or submitted work. To enhance the utility of these models, instructors deciding to share such data may introduce significant concerns about \textbf{\textit{privacy}} and \textbf{\textit{intellectual property}} rights \citep{kasneciChatGPTGoodOpportunities2023,yanPracticalEthicalChallenges2023}, complicating the adoption and effective use of these technologies in educational settings. 
Due to their design as conversational models, GenAI technologies and LLMs excel at delivering brief replies to single queries but struggle to effectively process a comprehensive set of batch tasks \citep{changSurveyEvaluationLarge2024}, such as offering personalized feedback to each student, limiting their use case for assisting with teaching at scale. Finally, the conversational format of these interfaces poses challenges in \textbf{\textit{validating their impact}} as they have limited capabilities for extensive data collection and for supporting rigorous research such as replication studies.

This paper investigates the potential benefits of transitioning from CUI-centric approaches, which directly use general-purpose GenAI systems, such as ChatGPT, to GenAI applications as custom-built software systems that make use of GenAI and feature user-centric interfaces, to mitigate the previously outlined challenges. These user-centric interfaces prioritize the educator's instructional goals, focus on workflows aimed to facilitate interacting with GenAI against these goals, and enable educators to oversee the whole output generation process while ensuring content accuracy. Accordingly, we first leverage cutting-edge research in the learning sciences, educational technology, learning analytics, and artificial intelligence to \textbf{\textit{develop a framework for pedagogically sound and ethically responsible integration of GenAI into educational tools}} and ecosystems. Our framework consists of two main components. The first component guides designing the core of a tool, which includes selecting educational tasks, applying pedagogical theories, setting criteria for evaluating GenAI outputs and data, and choosing GenAI models and prompting strategies. The second component focuses on designing user interactions with GenAI. This involves creating user interfaces and workflows, generating prompts, reviewing and assessing GenAI-generated content based on the criteria established in the first component, and producing the final content. A key aspect of our framework is that it empowers educators as disciplinary experts to oversee the entire process and to be able to ensure content accuracy with optimal use of time. 

We then demonstrate the application of our framework in action through the development of a ``GenAI Feedback Provisioning Copilot''. The goal of this tool is to aid instructors in creating personalized open-response feedback for students' graded assignments. Specifically, the Feedback Copilot tool takes as input a set of assessment tasks, sample solutions, and students' assignments that have been graded by the teaching team, as well as a set of standards for assessing the quality of the feedback that the tool generates. The tool then produces customized feedback for each student. Importantly, the feedback quality standards are used by the tool to autonomously identify instances where the generated feedback may not fully meet the established quality criteria. This feature enables instructors to specifically focus on and review such cases, ensuring that all feedback provided to students is of the highest quality. 
Finally, we report findings from an empirical evaluation of the tool in which we analyze the feedback generated for assignments submitted by 338 students.  We explore two variations of the tool to understand how additional role-based prompting and guidance for providing effective feedback affect the quality of the feedback produced. In addition, we explore the relationship between the quantitative grades of the students (categorised as high, medium and low) and the quality of the feedback assessed based on selected criteria (constructiveness, empathy, detail, actionability, and encouragement of self-reflection).

The study revealed that the more advanced variation of the Feedback Copilot tool, that included additional prompting and guidance for generating effective feedback, produced feedback surpassing the quality of feedback generated using a base version. We also found an association between lower assignment grades and lower-quality feedback. This underscores the importance of educator oversight in feedback generation, particularly for lower-performing students who necessitate more comprehensive and pedagogically sound feedback. These findings emphasize the appropriateness of the proposed framework through it's reference implementation, Feedback Copilot. Our discussion informs future directions for the design and use of GenAI in educational settings, highlighting its potential to enhance teaching and learning experiences. %

\section{Related work}

\subsection{The Application of Generative AI in Education}
In the past decade, there has been a surge in the application of AI in Higher Education. This includes the implementation of profiling and predictive modeling in universities, as well as a resurgence of adaptive and personalized learning systems  \citep{bondMetaSystematicReview2023}. More recently, Generative AI (GenAI), including Large Language Models, has been increasingly utilised and explored in various teaching and learning scenarios \citep{mazzulloLearningAnalyticsEra2023,denny2024computing}, especially in the development of course content and integrated in Intelligent Tutoring Systems and Adaptive Learning Systems.
Dickey and Bejarano \citeyearpar{dickeyModelIntegratingGenerative2023} proposed a framework for employing GenAI in course content development. They posited that GenAI could facilitate the creation of innovative content or the refurbishment of content from previous course iterations.
Afzaal et al. \citeyearpar{afzaalTransformerBasedApproachAutomatic2023} devised and evaluated a method for generating automatic exercises tailored to students' learning needs, utilising the transformer-based model T5. Moore et al. \citeyearpar{mooreAssessingQualityMultipleChoice2023} found that traditional rule-based methods could still outperform LLMs in certain tasks compared to LLMs. In contrast, Denny et al. found that the overall quality of educational content generated by an LLM was comparable to that of the content produced by students as part of learning activities \citeyearpar{dennyCanWeTrust2023}.
Pardos and Bhandari \citeyearpar{pardosLearningGainDifferences2023} compared students' reliance on hints generated by ChatGPT to hints formulated by Teaching Assistants (TAs) in the context of school maths tasks. The authors conducted a randomised between-subject study with two conditions. The results indicated that both hints generated by ChatGPT and those created by TAs led to positive learning outcomes. Khosravi et al. suggested that GenAI could be effectively used in learnersourcing systems \citeyearpar{khosraviLearnersourcingAgeAI2023}. For instance, GenAI might be leveraged to provide students with initial templates for the tasks they are required to author. In this case, they would not be required to create tasks from a blank page.

\subsection{The Foundations of Automated Feedback}
A substantial body of evidence suggests that feedback positively influences learners by aiding them in improving their learning strategies, leading to enhanced academic outcomes \citep{hattiePowerFeedback2007,wisniewskiPowerFeedbackRevisited2020}. However, the quality of the feedback has traditionally depended on the instructors' ability to compose such feedback, which is effort-intensive. Automated feedback systems aim to address this issue \citep{keuningSystematicLiteratureReview2019,cavalcantiAutomaticFeedbackOnline2021,deevaReviewAutomatedFeedback2021}.
There exists a substantial number of systematic literature reviews, each delving into a specific facet of automated feedback. A systematic review of the literature on automated feedback in online learning environments by Cavalcanti et al. \citeyearpar{cavalcantiAutomaticFeedbackOnline2021} posited that automated feedback significantly alleviates the workload of instructors when administering the course. Keuning et al. \citeyearpar{keuningSystematicLiteratureReview2019} confirmed that automated feedback systems in the context of introductory computer science courses (CS1) provide adequately accurate task feedback, which is however, typically quite uniform. A review by Maier and Klotz \citeyearpar{maierPersonalizedFeedbackDigital2022} suggested that current automated feedback systems lack personalization capabilities to account for students' unique backgrounds. One of the challenges with automated feedback systems is that they often necessitate extensive configuration to accommodate the course context and existing educational data, posing difficulties for educators to deploy them. Similar to automated feedback systems, LLMs hold the potential to be instrumental in scaffolding the generation of feedback at scale, while providing more benefits for personalization. 

\subsection{LLMs for Automated Feedback} The current use of LLMs in automated feedback is represented by two distinct approaches. The first approach utilises LLMs to obtain a representation of students' data in the form of embeddings \citep{berniusMachineLearningBased2022,linUsingLargeLanguage,gombertAutomatedAssessmentStudent2024}. The second approach leverages the generative capabilities of LLMs \citep{nguyenEvaluatingChatGPTDecimal2023,hanFABRICAutomatedScoring2023}. 
Bernius et al. \citeyearpar{berniusMachineLearningBased2022} proposed a machine learning approach to provide granular feedback for instructors grading students' tasks in the context of an introductory computer science (CS1) course. The authors utilised both traditional Natural Language Processing approaches and transformer models such as ELMo to represent students' solutions as embeddings, cluster these representations, and match them with existing solutions used as recommendations for the tutors. Similarly, a study by Gombert et al. \citeyearpar{gombertAutomatedAssessmentStudent2024} did not use LLMs to generate content and instead relied on LLMs to create embedding representations of the students' input data. The authors found that the approach had a positive effect on the students' perception of feedback. 
In contrast, Nguyen et al. \citeyearpar{nguyenEvaluatingChatGPTDecimal2023} explored the quality of LLMs' generative capabilities in providing feedback. Results indicated that the accuracy of such LLMs-generated feedback was generally aligned with that of the feedback formulated by tutors; however, such LLM-generated feedback contradicted the initially supplied instruction. Han et al. developed an automated feedback tool for English essay writing \citeyearpar{hanFABRICAutomatedScoring2023}. The resulting feedback was evaluated with both students and instructors; however, the differences between advanced and standard prompting pipelines were minor.
Overall, generative capabilities of LLMs show promising results, yet, the resulting feedback does not consistently adhere to the instructions embedded within the application. These issues could potentially be mitigated not only through technological advancements that improve the quality of LLM models but also at the user interface (UI) level.

\subsection{GenAI and LLM Interfaces for Guidance and Prompt Engineering Support} %
Interactions with GenAI, particularly with LLMs, are primarily defined by the UI paradigm of CUIs. From the user's perspective, most interactions with these models are realized through the formulation of user queries in natural language. 
However, recent research has unanimously found that users interacting with GenAI through CUIs commonly encounter two major challenges \citep{jiangGraphologueExploringLarge2023,suhSensecapeEnablingMultilevel2023} including an expertise barrier in prompt engineering and inability of CUIs to effectively guide users step-by-step in a comprehensive set of activities. 

\textbf{AI Literacy and Expertise in Prompt Engineering.}  To obtain the intended output, users are increasingly involved in formulating specific commands known as prompts. The purpose of these prompts is twofold: to instruct GenAI to generate more complex outputs, and to chain together intermediate results of GenAI call executions \citep{dangHowPromptOpportunities2022}. This process, known as prompt engineering, requires users to engage in the development, trial, and testing of different queries against the desired output. However, prompt engineering requires specific technical expertise, which is not universally available and impedes the widespread use of GenAI for a variety of scenarios.  Despite its apparent simplicity, recent research has revealed that non-experts often struggle to successfully formulate prompts that scaffold their use of GenAI \citep{zamfirescu-pereiraWhyJohnnyCan2023,kimUnderstandingUsersDissatisfaction2023}. This has led to a significant body of research focusing on simplifying the task of prompt optimisation for end-users, either by disseminating effective prompt techniques \citep{whitePromptPatternCatalog2023,cainPromptingChangeExploring2023} or by providing users with authoring tools and specialised environments assisting users in comparing how variations in prompts influence the final output \citep{dangHowPromptOpportunities2022,zamfirescu-pereiraWhyJohnnyCan2023,arawjoChainForgeOpensourceVisual2023,kimEvalLMInteractiveEvaluation2023}. 

\textbf{Excessive User Autonomy and Lack of User Guidance.} The introduction of GenAI presents a novel challenge for interface design \citep{WhyCodesigningAI}. As a response, numerous studies have explored the potential of alternative interface paradigms for GenAI to surpass CUIs \citep{labanChatExecutableVerifiable2018,PromptingDiscoveryFlexible,suhSensecapeEnablingMultilevel2023,angertSpellburstNodebasedInterface2023}. These interfaces offer more explicit guidance to end-users by scaffolding the domain-specific activities that users engage in when performing tasks using GenAI and LLMs.
For example, \citet{arawjoChainForgeOpensourceVisual2023,angertSpellburstNodebasedInterface2023} proposed interfaces for supporting programming and creative coding with LLMs, \citet{PromptingDiscoveryFlexible,huhGenAssistMakingImage2023,liu2022opal} designed UIs guiding users in activities such as creating concept art and images for news articles, scaffolding story writing \citep{mirowskiCoWritingScreenplaysTheatre2023,labanChatExecutableVerifiable2018} and general text writing \citep{suhStructuredGenerationExploration2023}. \citet{jiangGraphologueExploringLarge2023} designed UIs to support information-seeking and sensemaking tasks. Drawing on these emerging works, \citet{kim2023cells} formalised and proposed a set of scaffolding primitives for use by interface designers in creative tasks.
However, the guidance provided in these interfaces relates more to high-level tasks such as creative writing or programming activities, often during the exploratory phases of the task \citep{suhStructuredGenerationExploration2023,kim2023cells}. As such, the design principles formulated in these works are not easily transferable to the educational setting, which would have a rather specific set of requirements and activities to undertake the task. One of these requirements is the integration of institutional educational data to tailor the outputs of GenAI models. It is currently the responsibility of educators to determine which data can and cannot be used as input for GenAI \citep{kasneciChatGPTGoodOpportunities2023,yanPracticalEthicalChallenges2023}.

\textbf{Joint Inability of GenAI and CUIs to Support Effective Execution and Evaluation of a Comprehensive Set of Tasks.} LLMs are susceptible to hallucinations \citep{yeCognitiveMirageReview2023}. Furthermore, GenAI and LLMs do not support the execution of complex tasks \citep{changSurveyEvaluationLarge2024}, which would typically necessitate the user to engage in task decomposition themselves, prompting the model at each step to obtain intermediate results. This process would require the user to ensure the quality of the multiple generated outputs aligns with the task objectives. Providing a UI and backend to autonomously chain calls to an LLM opens up new possibilities for solving complex problems and automating educational workflows by leveraging the power of LLMs to handle tasks that are beyond the scope of a single call or message.  
For example, in contrast to engaging in iterative multi-turn conversational sessions with LLMs, tools implementing such an architecture would allow educators to create a set of assessment items by uploading their lecture notes to the application, choosing the desired number of items to generate, and consequently configuring the type of each assessment item and item's difficulty. 
Recent research has proposed several overarching mechanisms that could potentially mitigate these issues. \citet{huhGenAssistMakingImage2023} delved into the design of accessible UIs for interacting with GenAI for blind and low-vision participants. A comparable approach was presented by \citet{kimEvalLMInteractiveEvaluation2023}, who devised an interface to modify prompts based on user-defined evaluation criteria. The crux of the mechanism is to employ a less advanced GenAI model to ensure that the initial user requirements are met. In contrast, Gero et al. \citeyearpar{geroSupportingSensemakingLarge2024} explored how different UI structures, in conjunction with an innovative algorithm text-matching algorithm, could support users in comparing outputs at scale.

\subsection{Research Gap and Contribution} 

Recently, several scoped literature reviews have underscored the implications of GenAI use in teaching and learning scenarios. \citet{yanPracticalEthicalChallenges2023}, \citet{chiuSystematicLiteratureReview2023}, and  \citet{mazzulloLearningAnalyticsEra2023}   highlighted the potential of GenAI and LLMs for enhancing instructors' pedagogical practices and supporting professional development, alleviating instructors' workload, scaffolding instructors' ability to instruct. However, the authors also emphasised that these outcomes could only be achieved contingent on GenAI-enabled tools' design being closely aligned with the educational tasks instructors engage with. 
In order to enable these benefits of GenAI, the current study proposes a framework informing the user-centric design of GenAI applications. Our framework informs GenAI-enabled application design beyond the CUI-centric paradigm, lowering the expertise barrier to prompt engineering and providing step-by-step guidance to effectively aid educators in executing a wide range of educational tasks.

\section{Empowering Educators through a Generative AI Framework}
\label{sec:framework}

\begin{figure*}[h]
\makebox[\textwidth][c]{\includegraphics[width=1\textwidth]{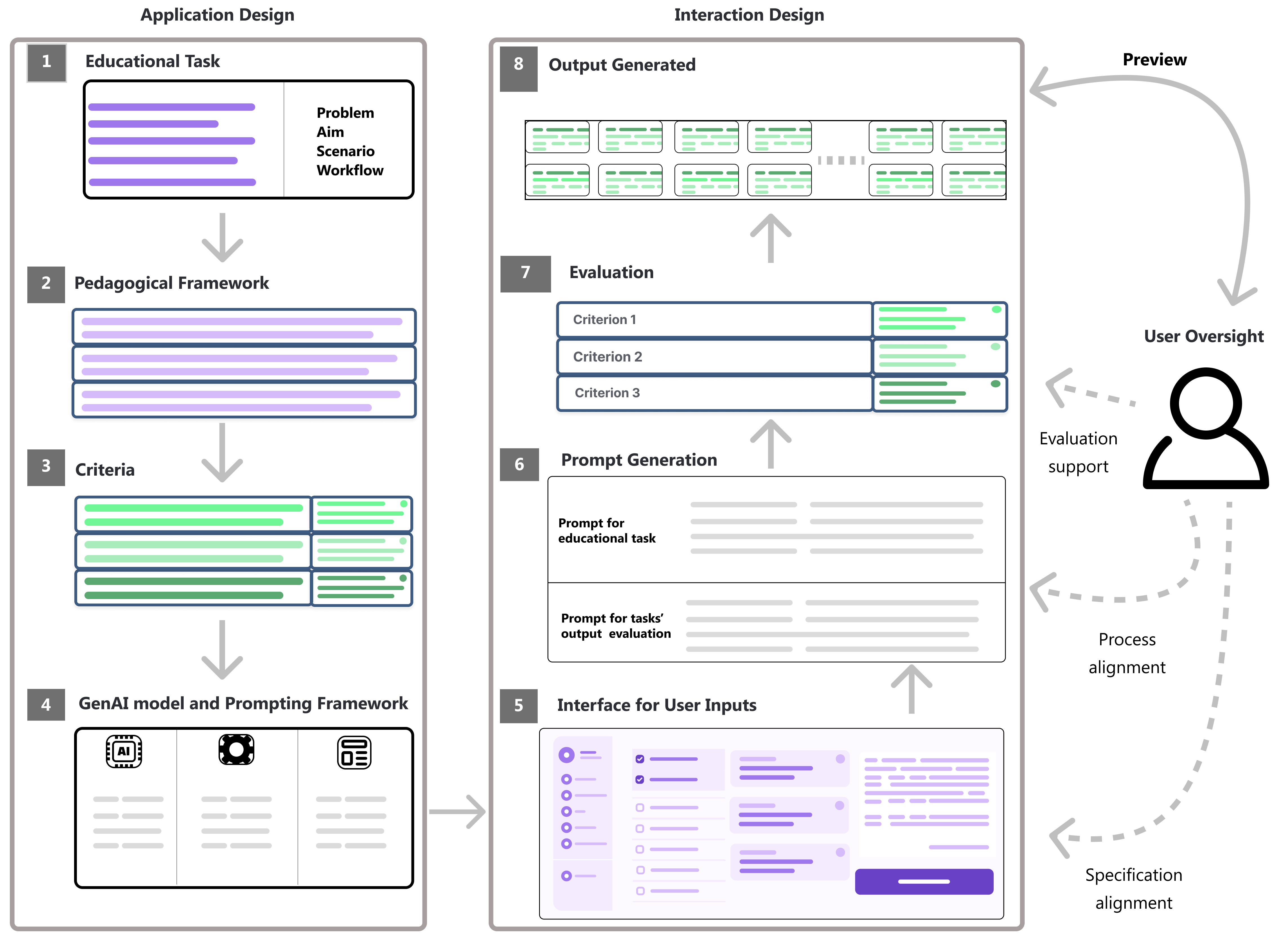}}
\caption{
The framework, formulated to aid the development of GenAI applications for educational tasks, consists of two components. 
\textbf{Component one} (Steps 1-4) guides the application design and high-level interface considerations, starting with the selection of the educational task and GenAI model, and ending with the creation of a user interface for GenAI model inputs. 
\textbf{Component two} (Steps 5-8) guides user interaction design, including steps for interactive intent alignment with the GenAI, prompt preview, active evaluation criteria selection, and GenAI model output preview with evaluation results.
}
\label{fig:framework}
\end{figure*}

This paper presents a design framework to facilitate the development of GenAI applications in education. 
We first reviewed works from the broader field of Human-Computer Interaction identifying reoccurring problems with how users interact with LLMs using CUI-based interfaces, as well as works describing approaches to overcome these problems. 
Then, we reviewed works describing issues with the application of GenAI for educational tasks. Based on these reviews, we propose a design framework which is aimed at resolving systemic gaps with GenAI and LLM interfaces applied to educational domains. 
The framework addresses previously identified gaps related to the CUI-centric design of GenAI and consists of two main components: the application design (Section \ref{sec:f-design}) and application interaction (Section~\ref{sec:f-interaction}) as outlined in Figure \ref{fig:framework}. 

\subsection{Application Design}
\label{sec:f-design}
The \textit{First Component} of the framework guides the application design, starting with the selection of the educational task and ending with data, and the GenAI model. This component, encompassing steps one to four, leads to the design of application interfaces. Designers choose a data source, GenAI class, prompting techniques, and consider the prompting pipeline, aiming to reduce the need for high AI literacy and prompt engineering expertise (\textbf{Gap 1}).
Designers then plan the inputs and sequence for task completion. This logic is later presented to the user, potentially as a sequence of interface views or forms, each prompting for information to complete the GenAI-scaffolded task. 
Designers should also consider existing institutional data sources for GenAI model input and their integration into the application following ethical guidelines.

\paragraph{1-Educational Task.} The first step in the framework entails identifying an educational task for a GenAI application. Designers and educational experts should consider tasks that are challenging and time-consuming for educators, yet offer substantial value. 
To be firm on the choice of an educational task for an application to support, three main considerations should be clarified a) what is the key \textit{problem(s)} with the current execution of the educational task, b) what are the applications' \textit{aims and objectives} to solve this problem(s), and c) how could users utilise the application for the educational task to overcome the problem(s), i.e., what are the \textit{scenarios} and \textit{workflows}.
For instance, GenAI can assist in tasks like providing personalized feedback or creating engaging content, freeing up educators' time. However, GenAI should not be used for high-stakes decisions impacting students' academic experiences, such as automatic grading. The choice of educational task influences the GenAI model and prompting templates. Tasks can include generating feedback, rubrics, learning content, discussion topics, and Q\&A materials.

\paragraph{2-Pedagogical Framework.} The next step is to select a pedagogical framework for the previously defined task.
This step is crucial as most researchers in AI in education (AIED), often with an engineering background, may not be familiar with various pedagogical approaches \citep{chiuSystematicLiteratureReview2023}. This lack of familiarity could lead to decisions influenced by market dynamics, such as over-optimism about new technologies or the use of redundant design solutions \citep{nieto_beyondDashboard_22}. They might also rely on outdated research or unproven concepts (e.g., learning styles), neglecting a diversity of teaching and learning approaches \citep{kirschnerLearnersReallyKnow2013,luckin_AIED_19}.
The choice of a pedagogical framework can be guided by literature or the preferences of the target audience, including learning designers and educators. During this step, it is important for designers, researchers, and educational experts to work together to select the most suitable feedback framework. This means they need to combine their expertise to ensure the chosen framework effectively addresses the educational needs and enhances the learning experience. This should lead to the identification of key theoretical constructs within the framework for later operationalisation.

\paragraph{3-Evaluation Criteria.} This step involves establishing evaluation criteria to validate the GenAI model's output.
Application designers need to decide \textit{who defines the criteria}, potentially consulting with users or educational experts to formulate task-specific criteria. Requirement elicitation and co-design approaches \citep{maldonado_DesigningTranslucent_20} can be used to ensure criteria are theoretically sound and practically relevant. Importantly, designers should exercise caution when using these approaches, considering the potential challenges and limitations, and opt-out if necessary.
Evaluation criteria might be formulated based on the educational task context (step one). Designers can consult the pedagogical framework and synthesize requirements for the final output. Iterative experimentation with GenAI can help formulate criteria ensuring these requirements are met in the final output.
Alternatively, designers could provide an authoring interface for users to define criteria during application run-time, as in \citep{kimEvalLMInteractiveEvaluation2023}.

\paragraph{4-Data, GenAI Model, and Prompting Template.} This step requires designers to consider existing institutional data sources, ensuring privacy and minimizing bias. Designers should engage with stakeholders and experts to incorporate these considerations into the application design.
Designers then select a GenAI or LLM model capable of generating the necessary output for the task, which could be structured or unstructured text, images, videos, or a combination of these modalities.
Given data availability and costs, designers may fine-tune an existing GenAI model. However, systematic prompting approaches like mutation prompts, emotional stimuli, and chain-of-thought are increasingly showing efficacy over fine-tuning \citep{noriCanGeneralistFoundation2023,fernandoPromptbreederSelfReferentialSelfImprovement2023}.
Upon model selection, designers determine the most effective technique for prompting LLMs, such as zero-shot, few-shot, or chain-of-thought \citep{schmidtCatalogingPromptPatterns}. Not all techniques may apply to certain tasks \citep{khattabDSPyCompilingDeclarative2023}. Depending on input complexity, designers may choose a specific prompting template to organize multiple input chunks for a single API call to the GenAI model.

\subsection{Interaction Design}
\label{sec:f-interaction}
The \textit{Second Component} outlines user interaction with the GenAI application. It includes steps for interactive intent alignment with GenAI, such as prompt preview, active evaluation criteria selection, and GenAI model output preview with evaluation results. These steps, five to eight, aim to guide users on required inputs and next actions.
These steps primarily inform interaction design and user experience, suggesting designers incorporate means of changing prompts or model parameters through the UI.
For instance, changing the diversity of the generated outputs, e.g., assessment items or feedback might be approached by both adjusting the model parameter values and/or the underlying prompt content.
However, designers may opt-out from exposing prompt changing functionality considering task constraints, prompt complexity, and user AI literacy (\textbf{Gap 1}).
Finally, designers should enable user oversight over generated output to account for hallucination risks in complex tasks (\textbf{Gap 3}). The steps are described below.

\paragraph{5-Interface Design.} This stage involves designing a UI to streamline input collection for the GenAI model from the user. The educational task, pedagogical framework, GenAI model, prompting framework, and evaluation criteria from steps 1-4 inform the required user input, leading to UIs unique for each task. For instance, a feedback generation UI would differ from a learning content creation UI due to input granularity and UI elements \citep{kim2023cells}. The goal is to create intuitive UIs that guide user input specification for the GenAI model, avoiding the complexity of explicit intent specification in natural language. Designers might allow open-ended responses, limit inputs with predefined options, or enable external resource uploads, such as rubrics or assignment descriptions.

\paragraph{6-Prompt Generation.} This step involves populating the prompting template(s) with user-specified inputs. Depending on decisions made in \textit{component one -- application design}, the application may need a sequence of prompts to generate the required output. Designers need to decide if users should be able to a) oversee these intermediate prompts and b) modify the prompts. Recent research highlights the challenges non-experts encounter when engaging in prompt engineering \citep{zamfirescu-pereiraWhyJohnnyCan2023}. Not all educational users may want to engage in complex multi-iterative prompt crafting and testing. Hence, designers could disable prompt modification functionality while revealing the prompts used by the application through the UI.
On-demand explanations could be an alternative to prompt modification. There is a growing body of evidence supporting the benefits of explainability in AIED \citep{khosravi_XAIed_22, longo2024explainable}, leading to improved technology adoption. Designers could reveal the underlying mechanisms of the GenAI and LLM models and/or incorporate explanations as to why a specific output is generated. It should be noted that explainability may be crucial for some tasks, e.g., automatic grading of students' assignments, and less so for others, e.g., generating discussion topics. Additionally, depending on advancements in GenAI and LLM explainability \citep{zhao_XGenAI_24}, revealing parts of the output generation process could improve transparency over the application's inner workings and foster trust. 

If designers allow users to modify prompts, they should consider how users would navigate this process. Designers could enable manual prompt editing or use innovative UIs like a node-based UI \citep{arawjoChainForgeOpensourceVisual2023} or draw from systems like Opal \citep{liu2022opal} or GenAssist \citep{huhGenAssistMakingImage2023}, which offer a more structured approach, substituting direct prompt editing with traditional UI elements.

\paragraph{7-Validation.} This step involves validating the GenAI output in relation to the educational task and user-specified inputs, based on the evaluation criteria outlined in \textit{step three}. Validation strategies could include delegating content evaluation to a subordinate LLM or applying a similar mechanism to a 5-15\% sample of the input \citep{gao_llmEval_24,kimEvalLMInteractiveEvaluation2023}.
Designers could use simple means like a traffic light metaphor for \textit{user oversight} or adopt novel techniques designed for LLMs \citep{geroSupportingSensemakingLarge2024}. For instance, evaluation criteria could be used to rank the output alignment with user requirements and educational theories, categorising output into three categories, each with varying degree of oversight importance. This stage equips users to validate the output and address critical outputs with low evaluation scores. Designers should provide means to modify outputs requiring user attention.

\paragraph{8-Output Generation and Spot-Checking.} This step necessitates the customisation of how the output is generated and displayed. The output form depends on the GenAI model type and anticipated output. Designers could permit various output forms, e.g. plain text, PDF, images, information visualisation, and ensure the final GenAI model output is conveniently presented alongside validation results \citep{kim2023cells,krishna2024genaudit}.
To allow users \textit{spot-checking} and make necessary adjustments to a particular generated artefact, which demands user attention, designers could present a view with evaluation scores alongside explanations for each score. To further ease the spot-checking, designers could include the initial inputs provided by the user to help them contrast the input and outputs. 
Designers need to decide how these criteria will be represented with the final output. Various representations could be employed \citep{kim2023cells}, such as colour-coded output based on each criterion. Alternatively, automatic scoring might be applied to the generated output, resulting in a numeric evaluation of alignment between the output and criteria, along with a textual explanation.
These recommendations allow users to make necessary adjustments for specific artefacts and proceed with the generation process for the rest of the artefacts.

\subsection{Providing GenAI Means to Support User Oversight}

GenAI application design poses unique challenges due to the unpredictable nature of AI outcomes and the complexity of AI control \citep{WhyCodesigningAI,terryAIAlignmentDesign2023,subramonyamBridgingGulfEnvisioning2023,glassmanDesigningInterfacesHumanComputer2023}. During oversight, users may find the output does not meet their initial requirements, is too diverse or unique, or is of poor quality only for a certain group of artefacts. Designers can facilitate alignment to help users achieve better tool outcomes \citep{terryAIAlignmentDesign2023}.
This can be achieved through \textit{specification}, \textit{process}, and \textit{evaluation} alignments. \textit{Specification alignment} (step five) allows users to modify GenAI model inputs. \textit{Process alignment} (step six) adjusts the GenAI model's input processing to achieve the desired outcome. \textit{Evaluation alignment} (steps five, seven, and eight) provides mechanisms for users to validate outputs \citep{terryAIAlignmentDesign2023}. These mechanisms collectively empower users to control the GenAI process during oversight. The correspondence between these mechanisms and steps is shown in Figure \ref{fig:framework}.

\section{Reference Model of the Framework}
\label{sec:results}

\begin{figure*}[h]
\includegraphics[width=1\textwidth,height=0.9\textheight,keepaspectratio]{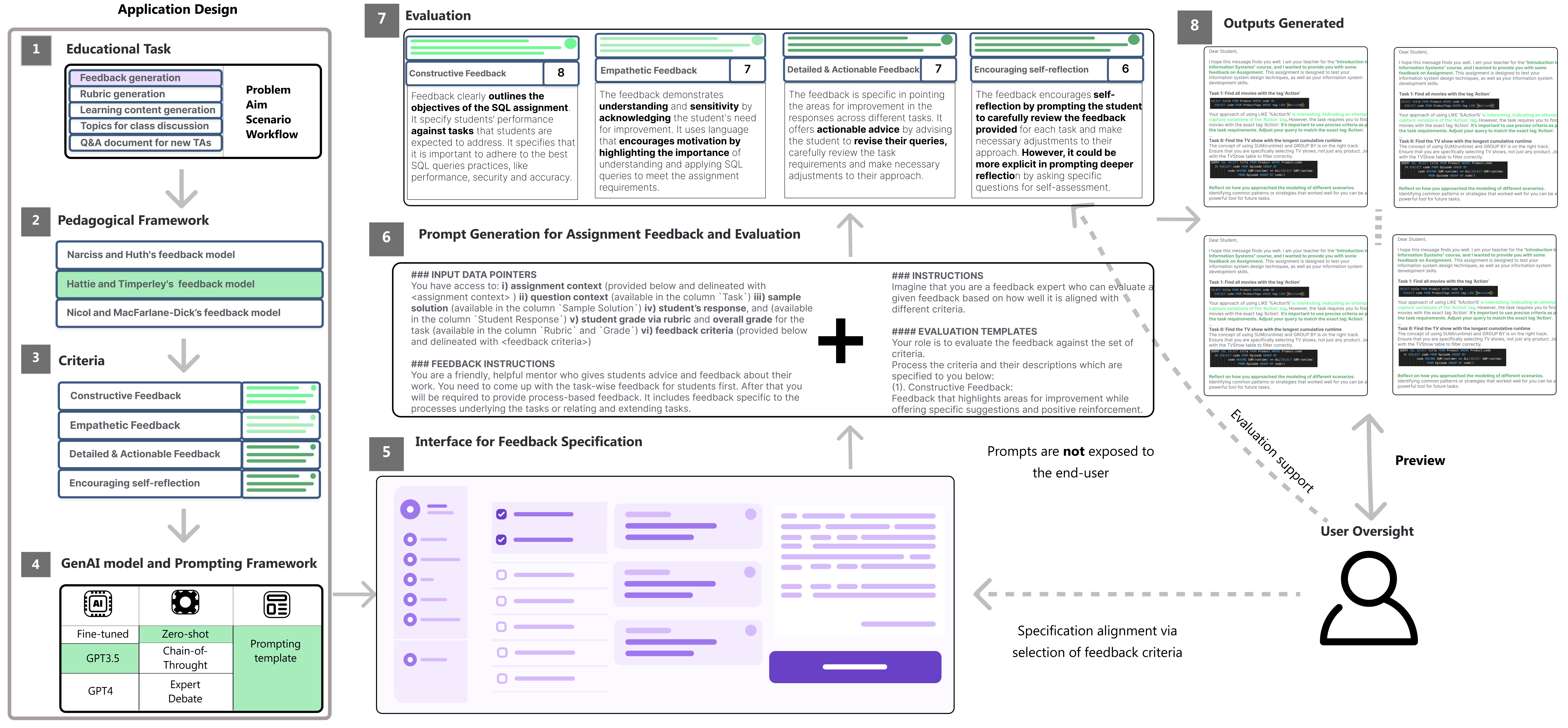}
\caption{This figure illustrates the framework's instantiation, demonstrating its application to inform Feedback Copilot's development. \textbf{Steps 1-4} demonstrate the potential alternatives for GenAI application development for feedback tasks. \textbf{Steps 5-8} show possible design decisions for UIs for instructor input, prompt pipeline interactions, evaluation pipeline, and generated feedback output.}
\label{fig:framework-instantiation}
\end{figure*}

It is crucial to equip instructors with the means to maintain control without a corresponding workload increase when using GenAI tools \citep{chiuSystematicLiteratureReview2023,mazzulloLearningAnalyticsEra2023}. This section introduces Feedback Copilot, a GenAI feedback tool designed using the previously described framework. The tool was designed to support university instructors teaching courses with a high student intake.
Figure \ref{fig:framework-instantiation} shows a reference model of the feedback-specific framework. 

\subsection{Feedback Copilot -- Application Design}
\paragraph{1-Feedback as an educational task.} Initially, a research team discussed different tasks and decided to focus on feedback as the educational task of interest. Several team members had substantial prior experience with feedback provision in various courses, thus, bringing their expertise as both educational experts and researchers.
The team of researchers discussed and formulated the \textit{problem} within existing feedback delivery systems as a generation of personalized feedback for students' assignments. The \textit{aim} of the envisaged tool was to enable the generation of personalized feedback for students' assignments while optimally allocating the instructor's time to ensure feedback quality. This led to the \textit{objective definitions}, which were 1) enabling oversight over the feedback generation with the possibility to spot-check some of it, 2) optimal instructors' time allocation during the oversight phase, and 3) easy integration with existing institution data sources. It was followed by outlining the main \textit{scenarios}, feedback generation and feedback evaluation, and \textit{workflows}, namely, how the instructor should be guided starting from input specification to an oversight phase. %

This led to defining Feedback Copilot as an aid for instructors in creating personalized feedback for students' graded assignments. 
Feedback Copilot takes as input a set of assessment tasks, sample solutions, students' assignments that have been graded by the teaching team, and a set of standards for assessing the quality of its generated feedback. The tool then generates customized feedback for each student. Importantly, it also automatically evaluates the quality of the feedback it generates, helping highlight where instructors might need to step in. This feature enables instructors to specifically focus on and review these cases, ensuring that all feedback provided to students is of the highest quality. %

\paragraph{2-Choosing a feedback framework according to the problem specifications and learning context.} We selected a feedback framework to guide Feedback Copilot's aspects: a) instructor inputs, b) the evaluation criteria for feedback, c) automation level, and d) integration with existing infrastructure. 
In this step, collaboration among designers, researchers, and educational experts is crucial for choosing the right feedback framework. Some frameworks guide the \textit{design of feedback}, i.e., student feedback message's structure and content (e.g., effective feedback practices \citep{nicolFormativeAssessmentSelf2006}). Others focus on \textit{delivery and automation aspects of feedback}, such as the timing and frequency, and feedback triggering events (e.g., Narciss and Huth's feedback model \citeyearpar{lipnevichReviewFeedbackModels2021} or Serral and Snoeck's automated feedback model \citep{serralConceptualFrameworkFeedback2016}). %

We used Serral and Snoeck's automated feedback framework \citep{serralConceptualFrameworkFeedback2016}, based on Hattie and Timperley's feedback model \citeyearpar{hattiePowerFeedback2007}. This framework was chosen for its ability to guide both \textit{feedback message design} and \textit{feedback automation} within existing LMS and educational tools.
During this step, the team of researchers discussed the \textit{feedback generation starting point}, \textit{feedback granularity}, and \textit{instructor involvement in feedback specification}.
We explored potential answers to the question, \textit{`What could be the starting point for instructors to engage with automatic feedback generation?'}. We considered starting with learner characteristics or feedback purpose, as suggested by Serral and Snoeck's framework \citeyearpar{serralConceptualFrameworkFeedback2016}.
The question of \textit{`What level of granularity should the feedback generated by Feedback Copilot target?'} was discussed by the team. Initially, we planned to offer task and assignment-level feedback but decided to focus solely on assignment-level feedback (an assignment is typically composed of several separate tasks).
The discussion then moved to the question \textit{`What degree of involvement should instructors have during the feedback specification stage?'} In line with Human-AI cooperation principles \citep{alfredo2024human}, we aimed to maintain instructor involvement while enabling feedback automation. We discussed various options to allow instructors to control feedback levels in the output, contrasting with existing tools that mainly provide task-level feedback. We expand on our design considerations in steps six through eight.

\paragraph{3-Feedback Evaluation Criteria.} This step, linked to \textbf{step two}, addresses challenges such as GenAI models' non-deterministic responses, hallucinations, and inconsistent adherence to prompts \citep{yeCognitiveMirageReview2023,zhangPreparingEducatorsStudents2023,kimEvalLMInteractiveEvaluation2023}. It is crucial to assure the instructor that output generated will be both meaningful and purpose-aligned. We approached it by establishing criteria for automatic GenAI output evaluation, equipping instructors with the means to validate the generated content.

Criteria formulation depends on the feedback framework chosen in \textbf{step two}. If the framework chosen for feedback does not provide criteria, designers and domain experts may use other frameworks or invite instructors and employ requirement elicitation techniques to inform criteria. The pedagogical framework we selected for step two, Serral and Snoeck's, does not explicitly provide criteria but is based on Hattie and Timperley's feedback framework \citeyearpar{hattiePowerFeedback2007}, which suggests effective feedback should answer: `Where am I going?', `How am I going?', and `Where to next?' \citep{hattiePowerFeedback2007}. These questions can inform evaluation criteria or be loosely rephrased as feedback criteria.

For the feedback evaluation criteria in step 3, inspired by Nicol and MacFarlane-Dick's seven principles for feedback that fosters self-regulated learning \citeyearpar{nicolFormativeAssessmentSelf2006}, we chose four evaluation criteria: i) constructive feedback, ii) empathetic feedback, iii) detailed and actionable feedback, and iv) feedback encouraging self-reflection and independence. We proceeded with predefined criteria, but designers might consider allowing instructors to define their own criteria via UIs, letting educational designers and instructors incorporate their own criteria into Feedback Copilot.

\paragraph{4-Data, GenAI Model, and Prompting Template.} This step involved the team discussing the choices of data sources at [Anonymised University], the GenAI model choice, a prompting framework, and suitable prompting techniques.
Firstly, we examined data from Gradescope and [Anonymised Data Source], both integrated with [Anonymised University]. Gradescope allows instructors to establish rubrics and grade assignments automatically or semi-automatically. For the reference implementation of Feedback Copilot, we used assignments submitted in a computer science course using Gradescope. During this step, we ensured that no sensitive student details were included in prompts sent to the LLM using pseudonymisation. [Anonymised Data Source], providing data on additional student learning activities, was initially considered but not included in the final version of Feedback Copilot as it did not influence the feedback.
Secondly, we chose to use the OpenAI `gpt-3.5-turbo-1106' model for feedback generation. As we only included text in the final feedback, a basic model without image generation capabilities was sufficient. During prototyping, it was clear that generating feedback for a single assignment required multiple GenAI API calls, leading us to choose a simpler, more cost-effective model.
Thirdly, a decision regarding the \textit{prompting techniques} to be used in the application needed to be made. We accommodated zero-shot prompting for feedback generation. Alternatively, tool designers can accommodate few-shot prompting or more advanced prompting techniques, such as chain-of-thought or chain-of-density, or opt-in for automatic prompt improvement techniques \citep{fernandoPromptbreederSelfReferentialSelfImprovement2023}. However, for certain educational tasks, these techniques might be unsuitable or challenging to apply. We considered a mixture of expert prompting techniques and found it beneficial, especially for providing criteria justifications. However, we decided to use zero-shot prompting both for feedback generation and evaluation due to cost constraints.

We employed a prompting template proposed by \citet{mollickAssigningAISeven2023}. The authors suggested a set of prompting templates and configuration steps necessary to create AI agents supporting various educational scenarios, such as AI as a coach, mentor, or teammate. The proposed prompting template comprises the following sections, which could be completed and fed into the GenAI model: \textit{role, goal, pedagogy, step-by-step instructions, personalization, and constraints}. However, adhering to all the steps outlined in the paper is time-consuming and might require instructors to be AI literate. Thus, the designers' aim in this step is to create a custom prompting template or to adapt an existing one such that it would incorporate all the input required for feedback generation. 
In our case, such inputs, which are required to be specified by the instructor, are assignment context, assignment tasks, and sample solutions for each student.

Our GenAI pipeline involves a three-step process. Firstly, we used a GenAI model specified above to generate task-wise feedback. Subsequently, we made a separate call to the GenAI model, which would combine task-wise feedback and synthesise it into assignment-wise feedback. Lastly, we used the GenAI model to provide an evaluation of the generated feedback against the set of criteria with a brief justification for each.

\subsection{Feedback Copilot -- Interaction Design}
\label{subsec:interface}

\begin{figure*}[htp!]
    \centering 
    \includegraphics[width=.7\textwidth,keepaspectratio]{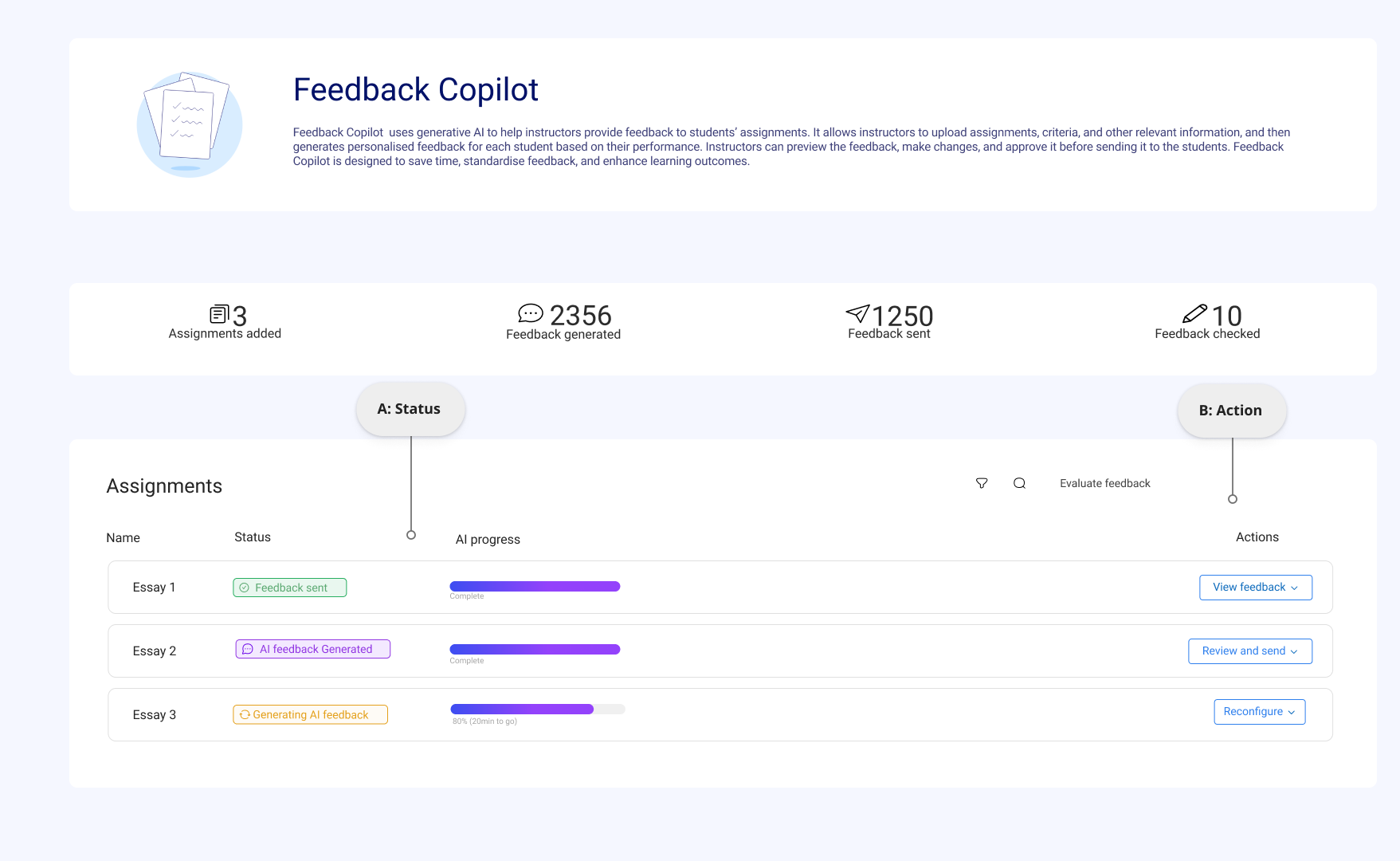}
    \caption{
     This figure provides an overview of the Feedback Copilot capabilities, enabling instructors to overview the status of the feedback generation (\textbf{A}) and action on the feedback (\textbf{B}). Each view is marked with a numbered overlay box, indicating a mapping to the corresponding step in the proposed framework.
    }
    \label{fig:interface-V1}
\end{figure*}

\begin{figure*}[htp!]
    \centering
    \includegraphics[width=.7\textwidth,keepaspectratio]{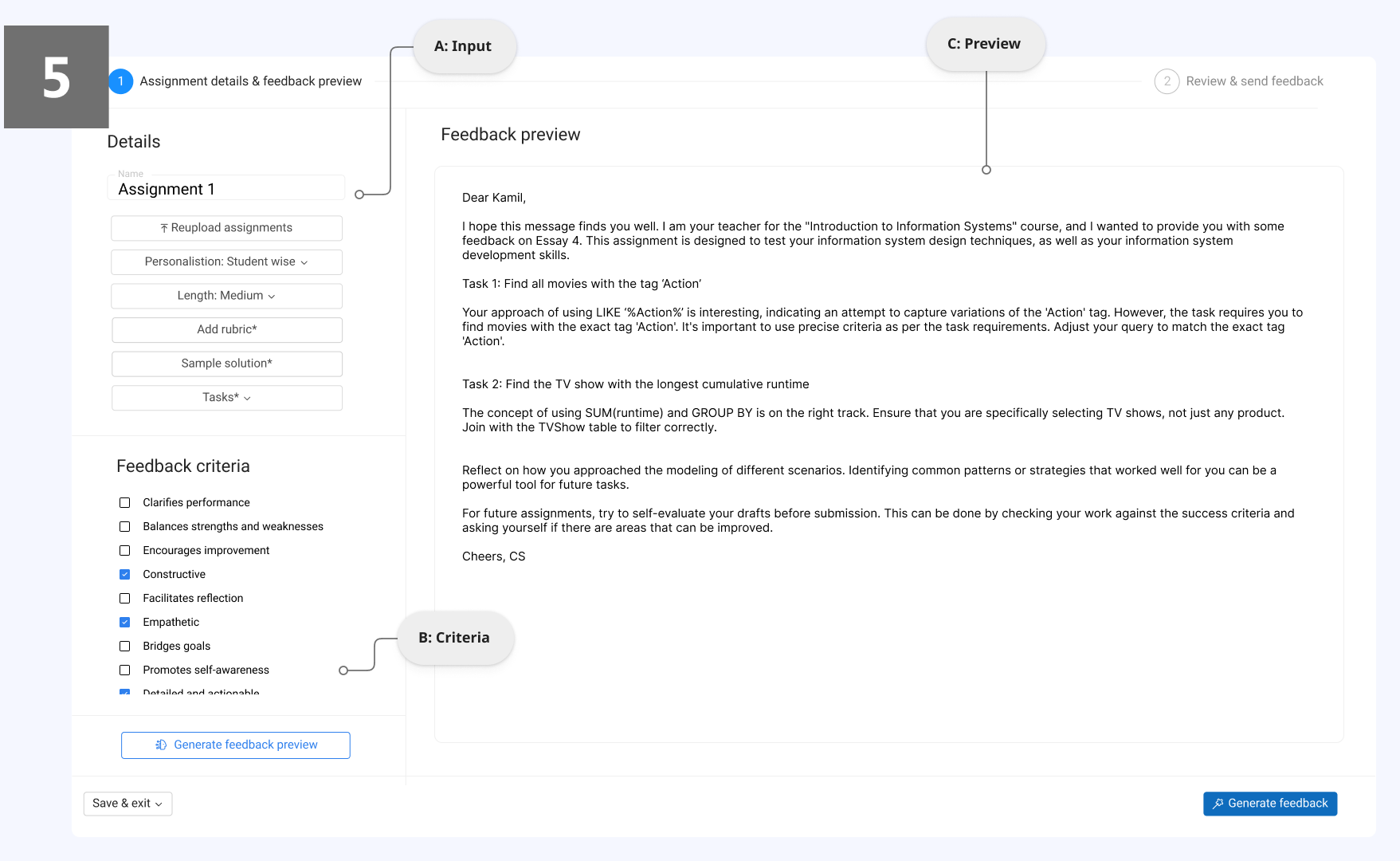} 
    \caption{
    The figure shows the Feedback Copilot interface. The instructor starts with specifying inputs to generate personalized feedback (\textbf{A}). Then, the instructor specifies feedback criteria (\textbf{B}). Feedback preview (\textbf{C}) provides a glimpse of a sample of generated feedback. Each view is numbered, mapping to the framework step.
    }
    \label{fig:interface-V2}
\end{figure*}

\begin{figure*}[htp!]
    \centering 
    \includegraphics[width=.7\textwidth,keepaspectratio]{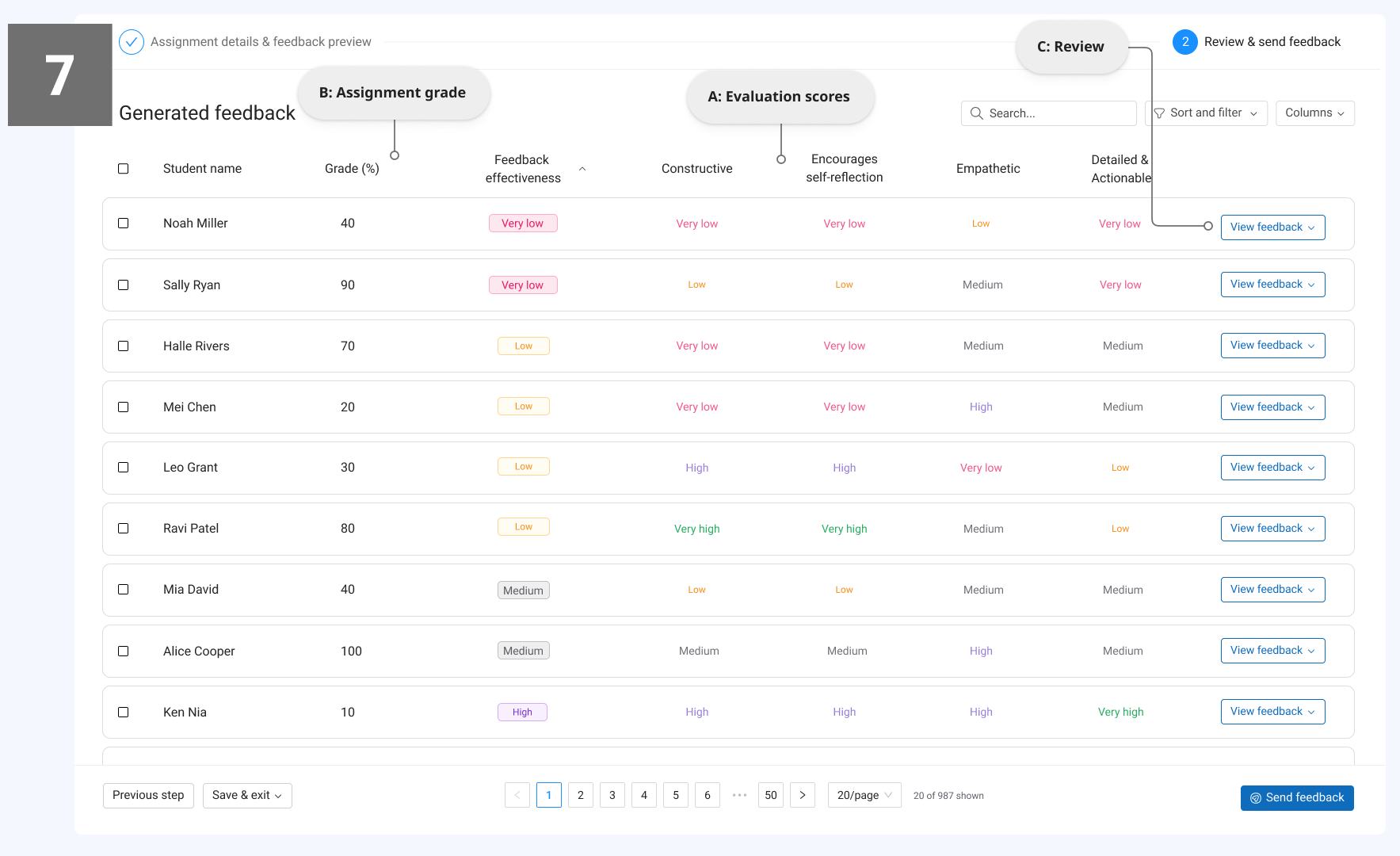}
    \caption{
     The figure provides an overview of the feedback generated for multiple students or student groups in the course, showing specific and overall feedback evaluation (\textbf{A}). 
     Additional information about students' backgrounds or assignment results could be included (\textbf{B}).
     It enables instructors to drill-down into individual overviews of the generated feedback that instructors can review (\textbf{C}). Each view is marked with a numbered overlay box, indicating a mapping to the corresponding step in the proposed framework.
    }
    \label{fig:interface-V3}
\end{figure*}

\begin{figure*}[htp!]
    \centering
    \includegraphics[width=.7\textwidth,keepaspectratio]{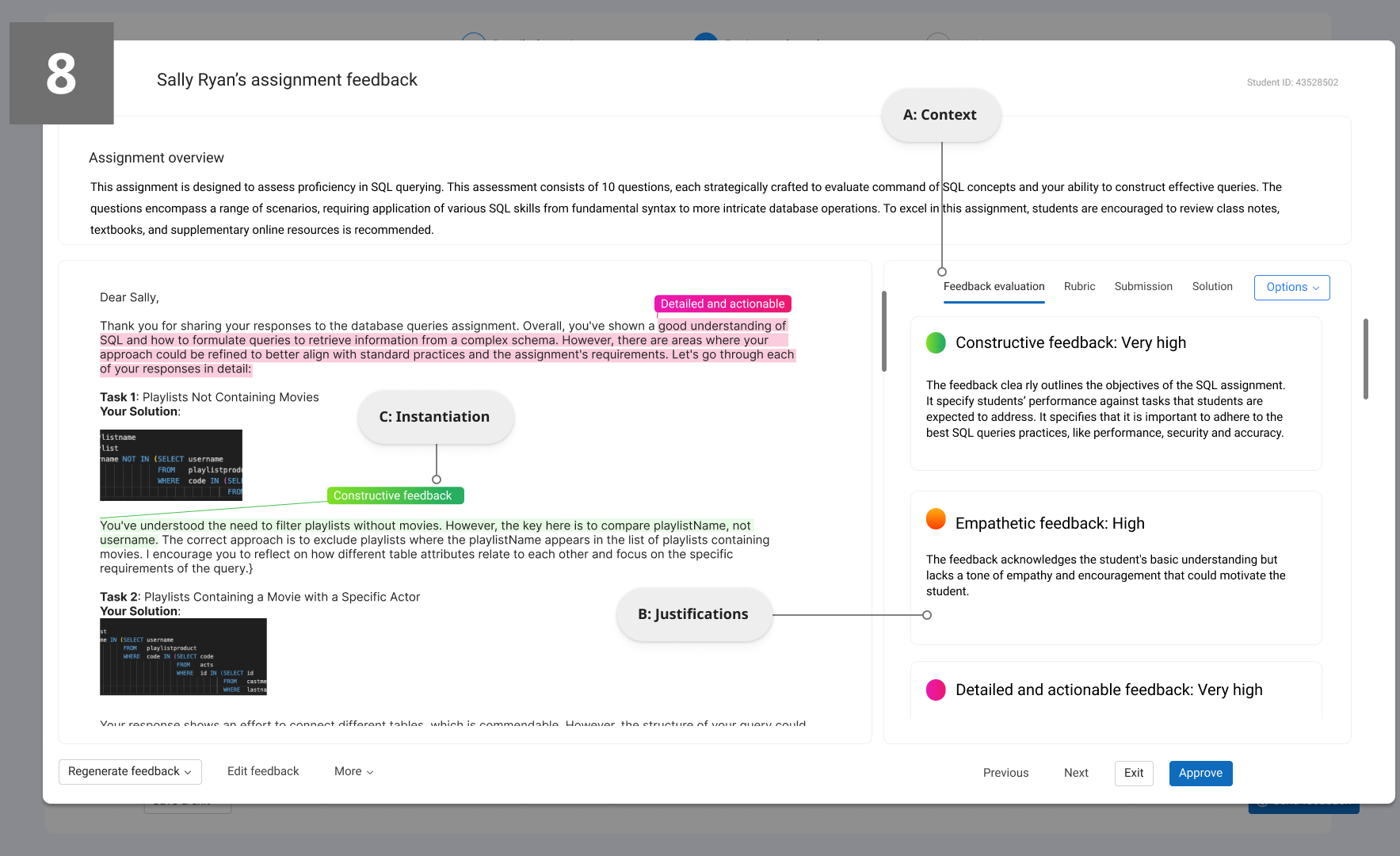} 
    \caption{
     This figure displays an inspection window for the generated feedback. Instructors can access the task-wise rubric and task description on demand (\textbf{A}), justification for each evaluation criterion (\textbf{B}), and criterion's instantiation with colour emphasis \textbf{C}). Each view is marked with a numbered overlay box, indicating a mapping to the corresponding step in the proposed framework.
    }
    \label{fig:interface-V4}
\end{figure*}

\begin{figure*}[htp!]
\centering
\includegraphics[width=0.7\textwidth,keepaspectratio]{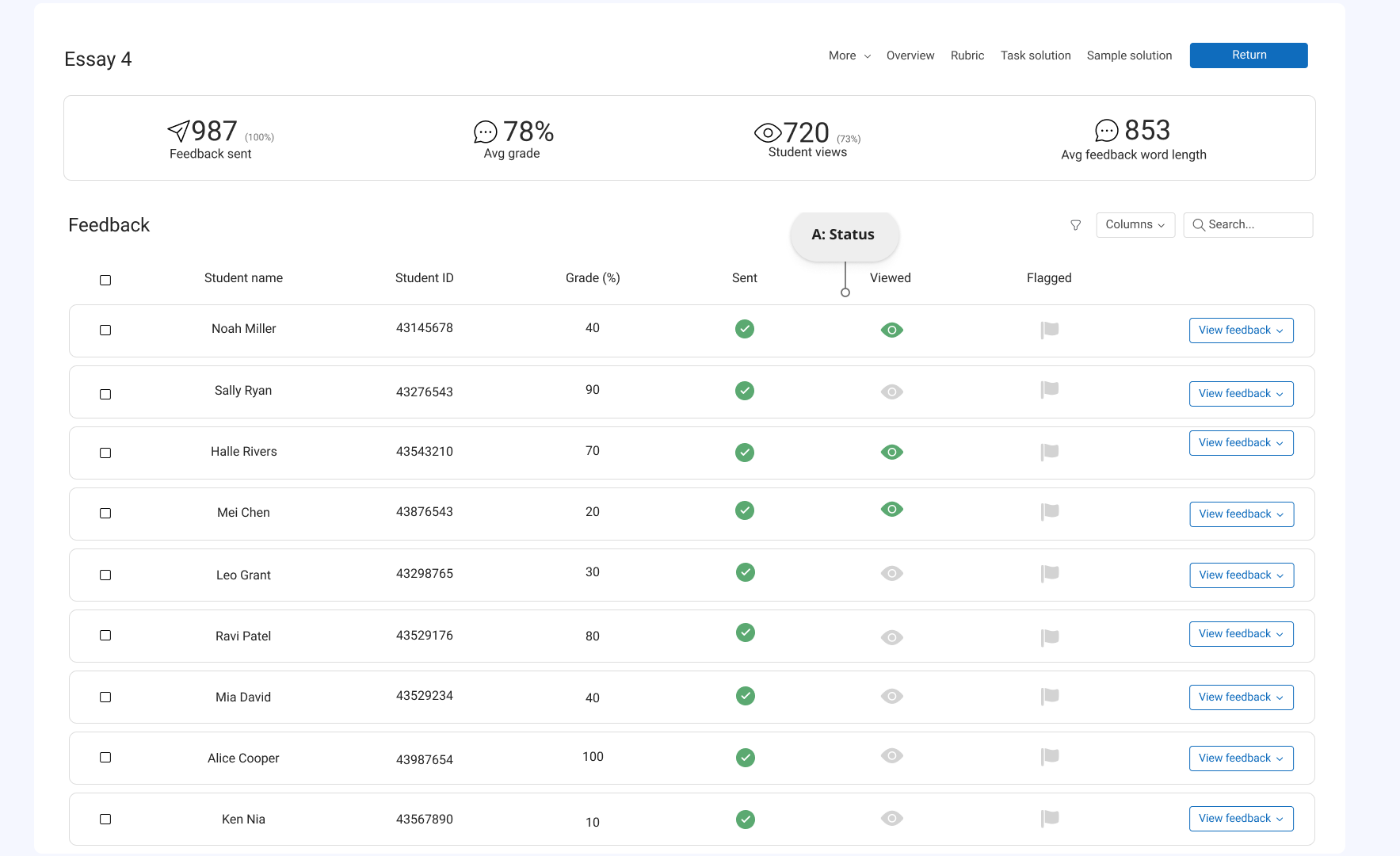}
\caption{
    This figure provides a comprehensive overview of the feedback delivered to students. It facilitates the review of whether students have viewed the delivered feedback (\textbf{A}) and includes an indicator for feedback that students have flagged as problematic. Student names and IDs shown here have been pseudonymised.
}
\label{fig:interface-V5}
\end{figure*}

\paragraph{5-Interface Design for Instructor Input.} This step involved designing an interface for an instructor to specify inputs for feedback generation. Initial designs were considered to include both open-ended inputs via text written in natural language and conventional UI elements. The final design adheres to a conventional linear structure, only permitting text file uploads for feedback generation, and various close-ended elements such as drop-downs, buttons, and toggles.
The GenAI-enabled tool for feedback generation, Feedback Copilot, is depicted in Figures~\ref{fig:interface-V1}, \ref{fig:interface-V2} and Figures~\ref{fig:interface-V3}, \ref{fig:interface-V4}, \ref{fig:interface-V5}. 
Instructors are expected to complete all the steps in views 5, 7, and 8, as numbered, respectively, on Figures \ref{fig:interface-V2}, \ref{fig:interface-V3} and \ref{fig:interface-V4}. 
The view depicted on top of Figure \ref{fig:interface-V1} presents an overview of all generated feedback.
The view depicted in Figure \ref{fig:interface-V2}-5b requires instructors to upload necessary documents for feedback generation, i.e., assignment documents, rubrics, students' submissions, and sample solutions, which are typically obtainable from Gradescope. Instructors then are required to choose evaluation criteria depicted in Figure \ref{fig:interface-V2}-5a), allowing the Feedback Copilot to generate a representative overview of the feedback for a specific student's assignment. 
Instructors can choose to generate individual feedback or feedback for the entire cohort, facilitated by separate learning analytics (LA). %
After providing all required inputs and specifying validation criteria, instructors can switch to an overview context (Figure \ref{fig:interface-V1}), which includes feedback being processed, feedback requiring review, and delivered feedback. This view allows instructors to monitor feedback generation time, identify feedback for review and dispatch, and examine previously delivered feedback.

\paragraph{6-Prompt Generation.} A pipeline was implemented to generate assignment-wide feedback, starting with task-wise feedback and then synthesising these into a single response. The prompts used in the pipeline are not exposed to the instructor, and there is no dedicated view for this step. This principle aligns with the challenges of prompt engineering, as it can be difficult for instructors to specify and modify their intents. Future versions of the Feedback Copilot may allow indirect prompt interactions, such as deciding on the inclusion of specific feedback components. This depends on data availability and the use of advanced prompt techniques.

\paragraph{7-Validation.} During this step, instructors are advised to select evaluation criteria for feedback. The GenAI model uses these criteria to evaluate and justify the feedback. The evaluation pipeline involves two calls to the GenAI model: one for scoring and explaining the feedback, and another for highlighting the satisfied criteria in the original feedback. The results are displayed on the UI in the following way.
A separate panel is designed for feedback validation, displaying scores and brief justifications for each criterion. Instructors can enable a colour emphasis overlay to see how the criteria are manifested in the feedback, and ask for explanations for each criterion instantiated in a student's assignment.
After setting up the feedback and evaluation criteria, instructors are directed to view 7 in Figure \ref{fig:interface-V3}. This view shows an overview of the feedback in relation to the criteria and allows instructors to oversee the feedback quality. It uses a traffic-light metaphor to categorise feedback into three levels based on the need for review before delivery to students: highly recommended for review, desirable for review, and ready for delivery to students, each necessitating a corresponding degree of instructor involvement.

\paragraph{8-Output Generation and Spot-Checking.} This step allows instructors to spot-check and correct feedback using two validation contexts: evaluation criteria and assignment context. 
The evaluation criteria align feedback with pedagogical recommendations from the literature. The assignment context enables factual issue detection and inconsistency resolution in the feedback.
A colour overlay representation of the evaluation criteria supports spot-checking. Each criterion's instantiation is emphasised with a corresponding colour in the feedback, with explanations for each criterion. Instructors can disable colour emphasis for some criteria to prevent visual clutter.
Instructors can verify feedback accuracy in terms of task specification adherence. The interface offers on-demand context windows for inspecting the complete assignment description, task-specific rubric, and sample reference solution. An in-tool editor allows feedback edits.  Instructors can also opt to regenerate the feedback.

After general validation (Figure \ref{fig:interface-V3}-7), instructors can spot-check individual student feedback. They can start with the lowest evaluated feedback by an inferior LLM. Clicking on a tile associated with individual feedback redirects to a drill-down view (Figure \ref{fig:interface-V4}-8). This view offers on-demand affordances for feedback checking and authoring editor capabilities for making edits prior to sending feedback to students.

Once feedback spot-checking and necessary edits are complete, instructors can mark the feedback as `approved' and proceed to the next feedback. After addressing instances, they can send the feedback to students (Figure \ref{fig:interface-V1}).
Finally, Figure \ref{fig:interface-V5} provides an overview of delivered feedback. Inspired by dialogic feedback \citep{yangFeedbackTriangleEnhancement2013}, it allows students to respond to received feedback. For instance, flagged feedback notifies the instructor of necessary corrections. %

\subsection{Providing GenAI Means to Support Instructor Oversight}

Figure~\ref{fig:framework-instantiation} schematically shows which features are included in the Feedback Copilot to support instructors making adjustments to feedback generation during the oversight process. These means are described below with examples of how alternative design choices might look.
\textit{Specification alignment} enables instructors to modify the inputs of a GenAI model, typically during the fifth step. 
In Feedback Copilot, explicit specification of the generated feedback by the instructor is currently present only by selecting from the list of evaluation criteria on which the resulting feedback should be based. Provision of inputs, such as student assignments, rubric criteria, and sample solutions, is not considered as a specification alignment.
Alternative ways to scaffold such alignment could be to provide a list of reference feedback examples, each with a different aim (e.g., to inform about weaknesses and strengths or to encourage students). Upon the instructor's selection, the GenAI model will generate feedback that mirrors the chosen example.
\textit{Process alignment} modifies how the GenAI model processes inputs to achieve the desired outcome, which occurs in the sixth step. The current implementation of the Feedback Copilot does not directly support process alignment (as described above). However, a variation of process alignment could include the introduction of a parameter that increases the variability of the generated feedback. This could be implemented as an interface element, such as a slider, which an instructor could use to increase the lexical variability of the feedback. 
\textit{Evaluation support} offers mechanisms for instructors to validate outputs. This begins when instructors define evaluation criteria in the fifth step and concludes during the seventh and eighth steps. The Feedback Copilot provides means for such alignment through both general validation (implemented via a traffic-light metaphor) and a mechanism for spot-checking, which allows for in-place modifications.  If instructors discover that the generated feedback either exhibits systematically poor quality or the feedback generated for certain student cohorts have poor quality, instructors could change the feedback evaluation criteria and regenerate the feedback.
The planned future iterations of the Feedback Copilot would include i) richer means to specify the feedback, and ii) means to diversify or homogenise feedback. 

\section{Evaluation of Assignment Feedback Generated via Feedback Copilot} 

The aim of the evaluation was twofold: firstly, to investigate the intrinsic qualities of the feedback generated by the Feedback Copilot tool, and secondly, to examine how the quality of this feedback varies depending on the performance of students on the assignments as measured by their assignment grades. The evaluation used assignment data from 338 students enrolled in an introductory undergraduate course on Relational Databases. The study also investigated the impact of two different configurations of the Feedback Copilot tool on feedback quality. Specifically, we assessed the feedback generated by a ``base'' version of the tool compared to an ``advanced'' version which augmented the inputs to the LLM with role-based prompting from the perspective of a mentor, detailed instructions for providing feedback, and the inclusion of explicit criteria for producing pedagogically effective feedback. Our analysis focused on i) the overall quality of the feedback, ii) specific dimensions of feedback quality, and iii) how equitably high-quality feedback is provided across students of varying achievement levels in the assignment. Guided by these considerations, our evaluation addressed the following research questions:

\begin{itemize}
    \item \textbf{RQ1}: How effective is `Feedback Copilot' at delivering high-quality feedback, and how does the quality of this feedback differ between the base and advanced versions of the tool?
    \item \textbf{RQ2}: In what ways do feedback quality dimensions such as constructiveness, detail and actionability, empathy, and encouragement of self-reflection and independence vary between the base version and the advanced version of the Feedback Copilot tool?
    \item \textbf{RQ3}: How is the quality of feedback from the base and advanced versions of the Feedback Copilot tool associated with students’ achievement levels on the assignments?
\end{itemize}

As such, the evaluation provides insights for edtech designers and educators regarding principles of designing more effective LLM-based tools for feedback and practical implications for teaching with LLM-based tools\footnote{This project is approved by anonymous ethics committee, Project ID: XXX}.

\subsection{Course Context}
We demonstrate how the feedback would look in the context of a Relational Databases course. This is a single-semester course and is typically taken by first-year undergraduate STEM students. This course aims to deliver foundational knowledge on the design and implementation of relational databases. The curriculum includes modules on data modelling, database design principles, the use of SQL for relational database queries, and the development of small-scale database applications utilising MySQL.

As part of this course, students were expected to complete three assignments and pass the final exam. For the purpose of this paper, we decided to focus on the feedback for the first assignment involving students working on SQL queries. The assignment consisted of five questions, each dedicated to writing an SQL query. This type of assignment typically includes a sample solution and a detailed rubric graded by TAs. Previously, LLMs showed promising results on generating and executing outputs involving programming and scripting languages \citep{becker_progHard_23}. Hence, we wanted to explore how detailed the feedback could be when the assignment involves an SQL query.
Typically, around five hundred students are enrolled in this course, which poses a considerable load on the teaching team to provide personalised feedback.

\subsection{Procedure and Data Collection}
\label{subsec:data}

\begin{figure}[!ht] 
    \centering
    \includegraphics[width=0.6 \textwidth]{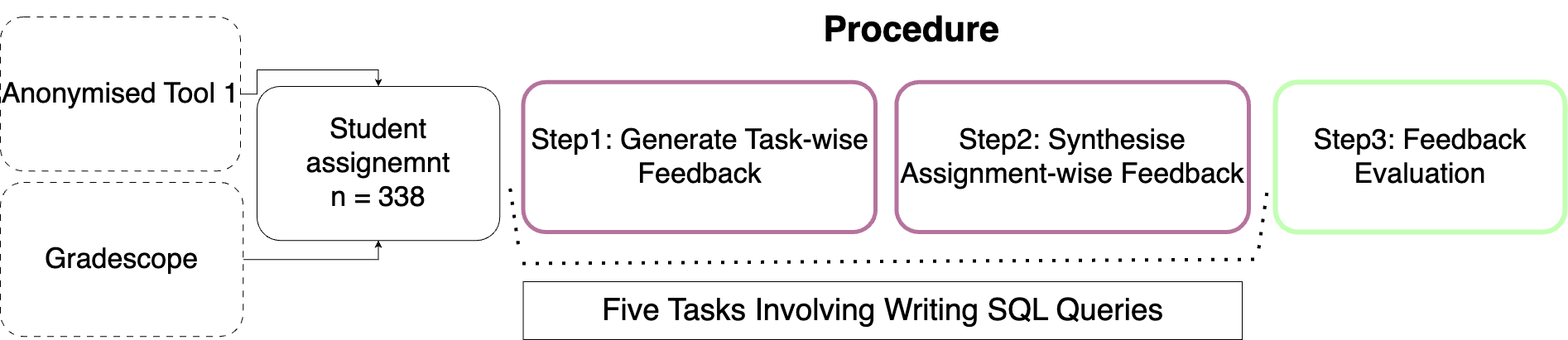}
    \caption{
        This procedure presents the sequence of steps used to generate and evaluate feedback using both variations of Feedback Copilot.
        }
    \label{fig:procedure}
\end{figure}

Personalised feedback generation used data extracted from two tools integrated into the Learning Management System (LMS), \textit{Anonymised Tool 1}, internally developed and deployed in the \textit{Anonymised University}, and Gradescope (see Figure~\ref{fig:procedure}). Data from the internal tool called \textit{Anonymised Tool 1} was used to utilise the students' engagement data regarding learning resources, such as accessing additional learning resources in the LMS, undertaking reading activities, and watching video lectures. Before generating the feedback, this data was used to cluster students into three groups depending on their engagement with different learning resources. Gradescope was used to extract the graded rubric for each task in the assignment.

To demonstrate the feasibility of our approach, we proceeded in the following way. We used the OpenAI API to i) generate feedback for 338 SQL assignments, and to ii) evaluate the resulting feedback. 
We then proceeded with the personalized feedback generation across two variations of the Feedback Copilot.
The \textit{base} variation of Feedback Copilot includes prompts requiring the following inputs to generate feedback to students: the task description and context, all student responses and corresponding numeric grades, the graded rubric, and a sample solution. The \textit{advanced} variation of Feedback Copilot extended the inputs for personalized feedback specification by including a) elements of role-based prompting for various educational scenarios \citep{mollickAssigningAISeven2023}, particularly indicating that the model should generate feedback as if it was written by the mentor of the course, b) detailed instructions on how to provide assignment feedback, and c) inclusion of the criteria of the pedagogically effective feedback as suggested in \citep{nicolFormativeAssessmentSelf2006}. 
The feedback generated using both Feedback Copilot variations, base and advanced, utilized the \textit{gpt-3.5-turbo-1106} OpenAI model. 
The underlying procedure to generate feedback consisted of two main steps (see Figure~\ref{fig:procedure}). Firstly, OpenAI API was used to generate feedback for each task. The total assignment consisted of five tasks, each of them requiring writing an SQL query. The second call to the API synthesised the task-wise feedback together. For demonstration purposes, we only used a subsample of five tasks which had the highest variation in student grades. 

To compare the quality of feedback resulting from two variations, we used \textit{evaluation} prompt (see Figure~\ref{fig:procedure}). This prompt was used to score the generated feedback based on four selected criteria of pedagogically effective feedback (suggested in \citep{nicolFormativeAssessmentSelf2006}) and to provide justifications for the resulting scores. Feedback was scored based on the following criteria: i) constructive feedback, ii) empathetic feedback, iii) detailed and actionable feedback, and iv) feedback encourages self-reflection and independence. We asked the model to provide scores on a range from zero to ten, which correspondingly maps to `not satisfying the criterion at all` to  `perfectly satisfied the criterion'. 
Table \ref{app:prompts} (in Appendix) contains prompt templates we used for the generation and evaluation of the feedback. Task descriptions and sample solutions are provided in Table \ref{tab:task-context}. Table \ref{tab:eval-prompt} includes a prompt template for scoring feedback and providing justifications. 

This was followed by an exploration of feedback generated via the proposed tool, Feedback Copilot. 
Particularly, we compared scores for feedback generated using two variations of the Feedback Copilot described above. The key motivation of this analysis was to explore that for students who would need high-quality feedback, e.g., students who received low and medium grades for the assignment, the model would generate feedback aligned with the criteria of pedagogically effective feedback. %
In the next subsection, we will refer to the Feedback Copilot variations as tool variations.%

\subsection{Data Analysis}
To answer question \textbf{RQ1} we used a one-way ANOVA model. The tool variation was included as an independent categorical variable (IV), and the average score across four criteria for feedback quality (described above in Section~\ref{subsec:data}) was used as a dependent variable (DV). 

To answer question \textbf{RQ2}, we used a MANOVA model. To explore the relationship between the tool variation and individually generated feedback quality as evaluated by the LLM model, we used a two-way MANOVA. Four DV variables were used in MANOVA, which were four feedback criteria: i) feedback constructiveness, ii) feedback empathy, iii) feedback being detailed and actionable, and iv) feedback encouraging self-reflection and independence. When conducting MANOVA, we used four tests, Pillai's trace, Wilks' lambda, Hotelling's trace, and Roy's largest root \citep{field2017discovering}, given that they have different robustness. 
The Royston test for checking multivariate normality was used and no violation of the assumption was detected for either the base Feedback Copilot ($H = 302.5694$, $p = 0$) or the advanced Feedback Copilot ($H = 322.3604$, $p = 0$). We used Box's M test to check the homogeneity of covariance assumption, which was satisfied ($x^2(10) = 87.331$, $p = 0$). We found no extreme cases of multicollinearity, with the largest Pearson correlation not exceeding $0.65$.

To answer \textbf{RQ3}, we used a two-way ANOVA model. In addition to the tool variation, we added a second categorical IV, which is students' achievement. We used the 33rd and 66th percentiles to divide students' achievement on the assignment into three levels, which are low, medium, and high achievement on the whole assignment. Before performing each ANOVA, we examined the data to ensure that the underlying assumptions were met \citep{tabachnick2013using}. 
Residuals were visually inspected and the departure from the normal distribution was only mild for both models. The homogeneity of variance was checked with Levene's Test and no violation was found in either case ($F(1, 674) = 1.7044$, $p = 0.1922$ for the model used in \textbf{RQ1} and for base Feedback Copilot  ($F(2, 335) = 0.164$, $p = 0.849$) and for the advanced Feedback Copilot ($F(2, 335) = 0.283$, $p =0.754$) for the model used in \textbf{RQ3}). Since we had an imbalance in the distribution of students' achievement across conditions, Type II ANOVA Tables were used in reporting \citep{field2017discovering}.

Both ANOVA and MANOVA were followed by a post-hoc pairwise analysis using the R package `emmeans' \citep{lenthEmmeans2022}. Bonferroni correction was used to adjust for multiple comparisons. Effect sizes for $Eta^2 partial$ were labelled following Field's \citeyearpar{field2017discovering} guidelines ($ES < 0.01$ -- very small, $0.01 <= ES < 0.06$ -- small, $0.06 <= ES < 0.14$ -- medium, $ES >= 0.14$ -- large). Cohen's d was used for post-hoc effect sizes, ($d <= 0.2$ -- small, $0.2 < d <= 0.5$ -- medium, and $d >= 0.8$ -- large).

\subsection{Results} 

\begin{figure}[!ht] 
    \centering
    \includegraphics[width=0.6 \textwidth]{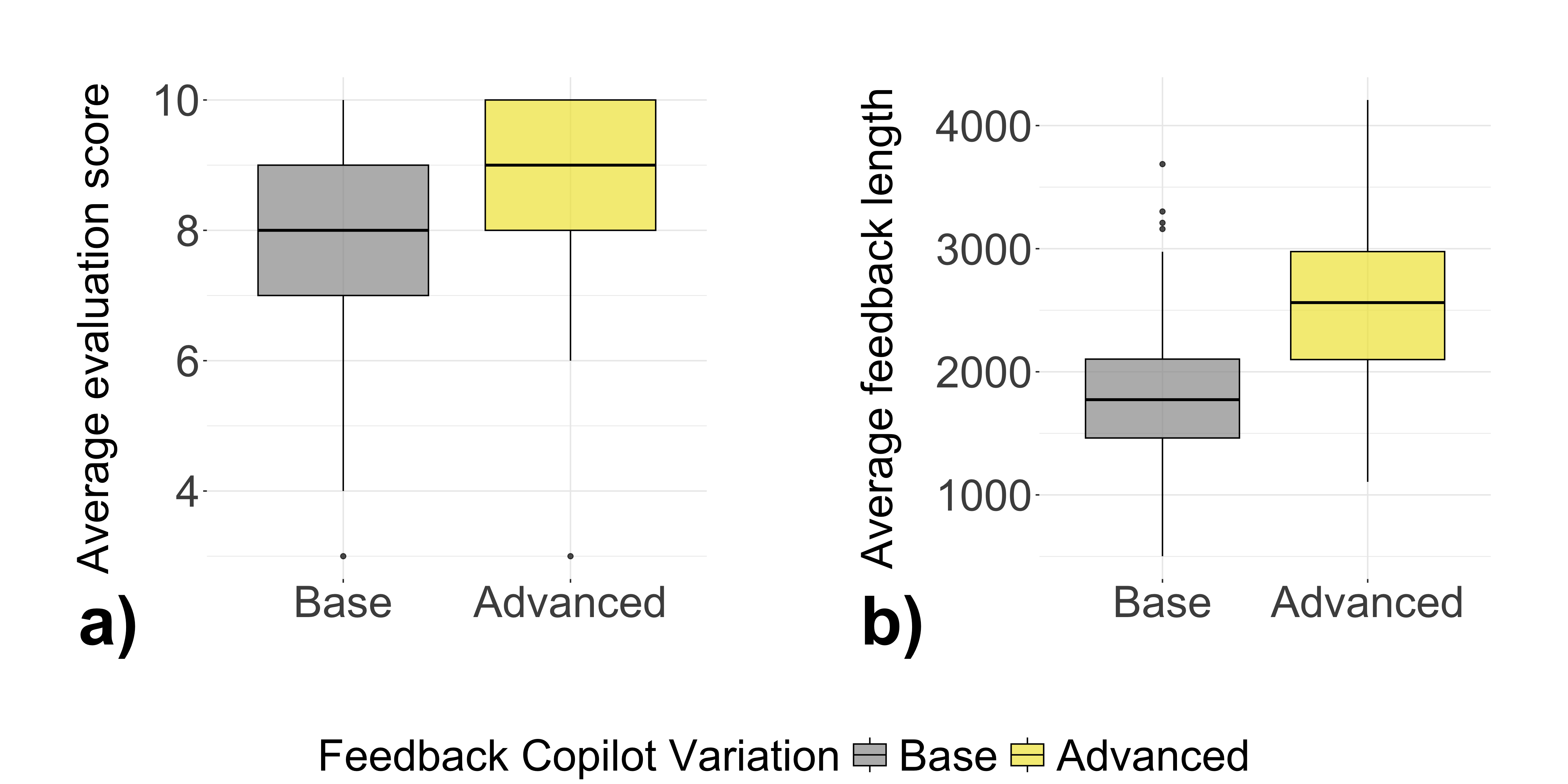}
    \caption{
        The subfigure \textbf{a)} presents the results of a one-way ANOVA. 
        It compares evaluation scores for feedback generated using base and advanced tool variation and Feedback Copilot.
        The y-axis represents the average evaluation score for the generated feedback.
        The subfigure \textbf{b)} represents the feedback length depending on which tool variation was used.
        The y-axis represents the number of symbols in the resulting feedback.
        }
    \label{fig:rq1}
\end{figure}

\paragraph{RQ1. Overall feedback quality.} The main effect of tool variation was statistically significant and large ($F(1, 674) = 260.49$, $p < .001$; $Eta2 = 0.28$, $95\%$ $CI [0.23, 1.00]$). Particularly, feedback generated with the base Feedback Copilot achieved an average feedback quality of $M\_base = 7.56$, while the average feedback score for one generated using Feedback Copilot achieved a score of $M\_advanced = 8.72$. Results are visually presented in Figure \ref{fig:rq1}~a).

\paragraph{RQ2. Individual feedback quality criteria.} The summary statistics from the MANOVA analysis are presented in Table \ref{tab:manova-table}. All four tests reached the criterion of significance of $0.05$. Results indicated that the tool variation had a statistically significant effect on the evaluation criteria combined. Following this, we conducted follow-up one-way ANOVA tests to analyse whether the effect of tool variation is achieved within each evaluation criterion.

\begin{table*}[!h] 
\centering
\caption{MANOVA test results for differences between tool variation's effect on the evaluation criteria scores. \label{tab:manova-table}}
\begin{tabular}[t]{llrrrr}
\toprule
Effect & Test & Value & F & Error df & P value\\
\midrule
 & pillai & 0.47 & 151.1063 & 671 & 0\\
\cmidrule{2-6}
 & wilks & 0.53 & 151.1063 & 671 & 0\\
\cmidrule{2-6}
 & hl & 0.90 & 151.1063 & 671 & 0\\
\cmidrule{2-6}
\multirow{-4}{*}{\raggedright\arraybackslash tool variation} & roy & 0.90 & 151.1063 & 671 & 0\\
\bottomrule
\end{tabular}
\end{table*}

\begin{figure}[!h]
    \centering
    \includegraphics[width=.7\textwidth,keepaspectratio]{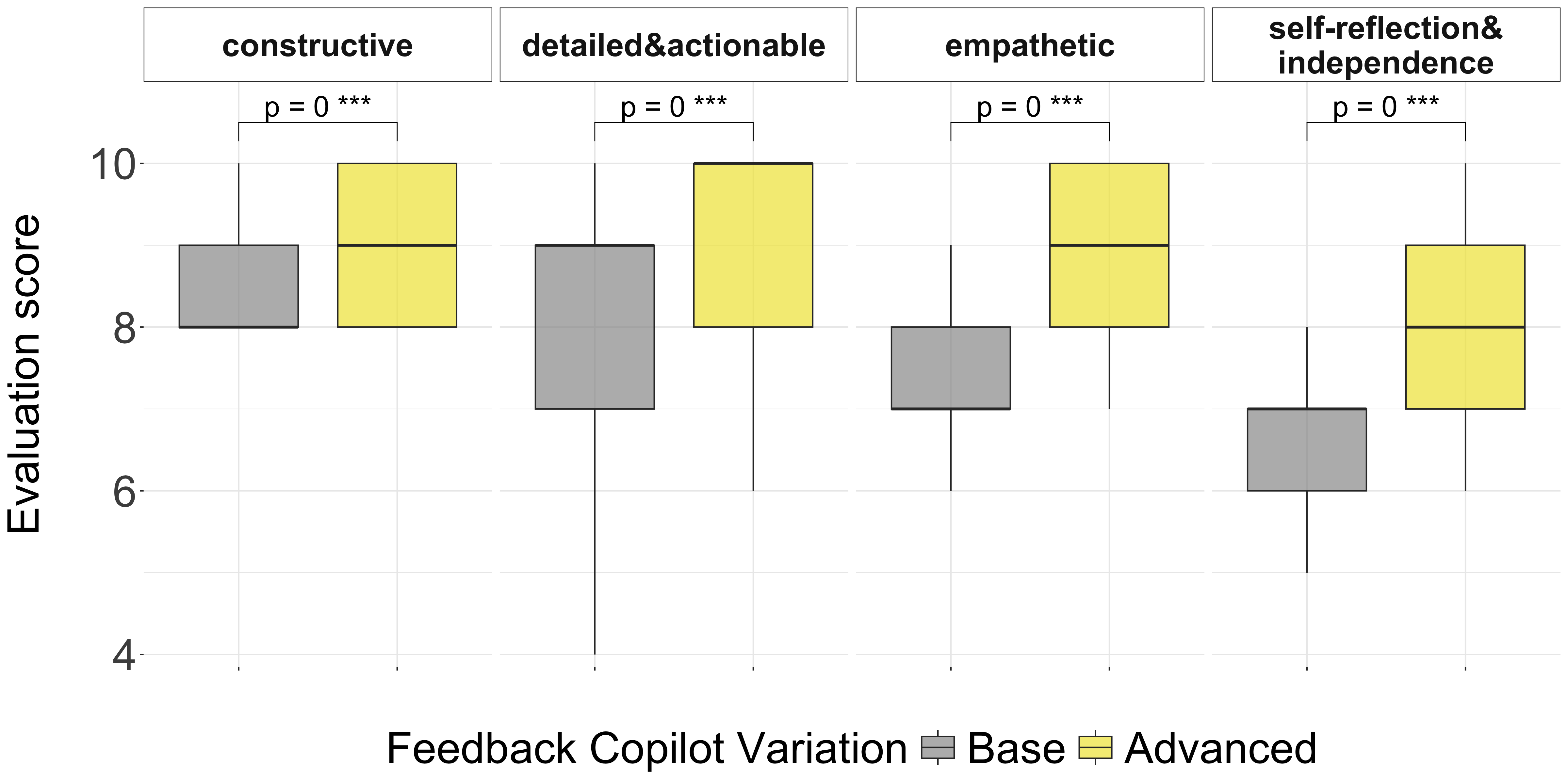}
    \caption{
        Results of the four ANOVAs following MANOVA, illustrating the distribution of feedback evaluation for each criterion depending on the tool variation used.
        The y-axis represents the evaluation score for each criterion.
        }
    \label{fig:rq2}
\end{figure}

Overall, all follow-up one-way ANOVAs revealed that the effect of tool variation on the score that the generated feedback would achieve was statistically significant for all of the evaluation criteria. All except for two criteria, `constructive feedback` and `detailed and actionable feedback`, showed a large effect size of using the advanced version of Feedback Copilot on the feedback quality. 
Particularly, in the case of the \textit{`constructive feedback'} criterion, a one-way follow-up ANOVA suggested that the main effect of tool variation was statistically significant and medium ($F(1, 674) = 72.33$, $p < .001$; $Eta2 = 0.10$, $95\%$ $CI [0.06, 1.00]$). For the criterion \textit{`empathetic feedback'}, a one-way follow-up ANOVA indicated that the main effect of tool variation was statistically significant and large ($F(1, 674) = 432.21$, $p < .001$; $Eta2 = 0.39$, $95\%$ $CI [0.35, 1.00]$). A one-way follow-up ANOVA suggested that the main effect of tool variation was statistically significant and small for the criterion \textit{`detailed and actionable feedback'} ($F(1, 674) = 19.05$, $p < .001$; $Eta2 = 0.03$, $95\%$ $CI [0.01, 1.00]$). A one-way follow-up ANOVA for the criterion \textit{`encouraging self-reflection and independence'} suggested that the main effect of tool variation was statistically significant and large ($F(1, 674) = 281.11$, $p < .001$; $Eta2 = 0.29$, $95\%$ $CI [0.25, 1.00]$)).
These results are visually presented in Figure \ref{fig:rq2}.

\paragraph{RQ3. Students' achievement and the overall feedback quality.} 
\begin{figure}[htp!]
    \centering
    \includegraphics[width=.8\textwidth,height=0.9\textheight,keepaspectratio]{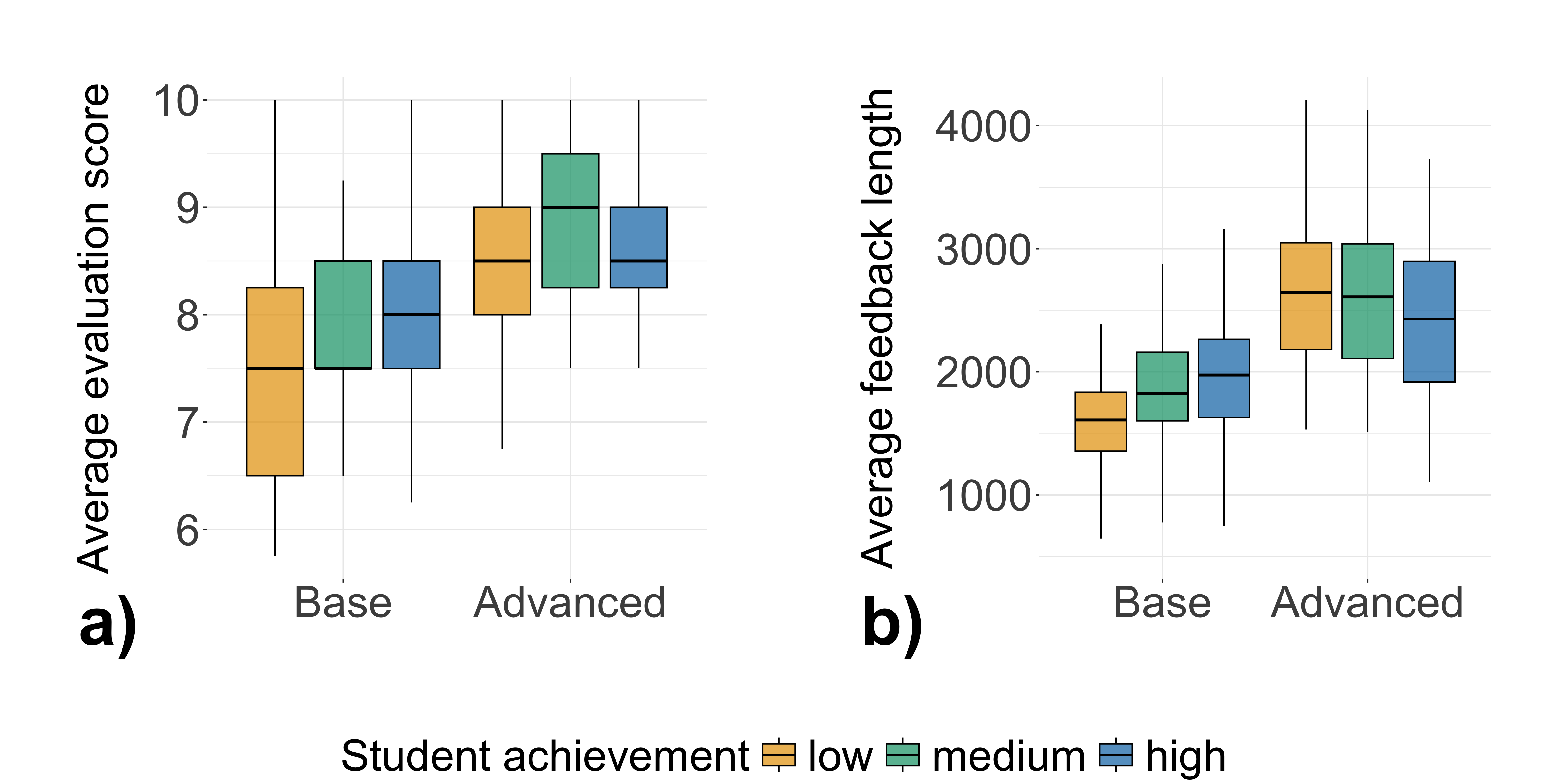}
    \caption{
        The subfigure \textbf{a)} presents the results of a two-way ANOVA. 
        The y-axis represents the average evaluation score for the generated feedback.
        It compares evaluation scores for feedback generated using a base and advanced Feedback Copilot variations for students who achieved low, medium, and high scores for the assignment.
        The subfigure \textbf{b)} represents the feedback length depending on which tool variation was used and students' achievement.
        The y-axis represents the number of symbols in the resulting feedback.
        }
    \label{fig:rq3}
\end{figure}

The main effect of tool variation was statistically significant and large ($F(1, 672) = 268.91$, $p < .001$; $Eta2 (partial) = 0.29$, $95\%$ $CI [0.24, 1.00]$). 
The main effect of students' achievement was statistically significant and small ($F(2, 672) = 11.89$, $p < .001$; $Eta2 (partial) = 0.03$, $95\%$ $CI [0.01, 1.00]$). 

Post-hoc analysis using adjusted pairwise comparisons revealed that there are statistically significant differences between the evaluation of the feedback generated for students with low achievement compared to students with medium achievement ($M\_low = 8.06$, $M\_medium = 8.41$, $t(672)=-4.8$, $p<0$, $d=-0.441$, $95\%$ $CI [-0.62, -0.26]$).
Significant differences were also identified between feedback evaluation for students who had low achievement compared to students who had high achievement ($M\_low = 8.06$, $M\_high = 8.27$, $t(672)=-2.7$, $p<0.006$, $d=-0.27$, $95\%$ $CI [-0.46, -0.08]$).
Figure \ref{fig:rq3}~a) presents the distribution of average evaluation scores depending on students' achievement on the assignment and tool variation.

We conducted an analysis of the feedback length depending on the tool variation and students' grades on the assignment. Figure \ref{fig:rq3}~b) indicated that feedback generated via the advanced Feedback Copilot had a longer length compared to the feedback received through its base variation. Notably, for the advanced version of the tool, the feedback generated for students who achieved lower grades on the assignment had a higher median length in comparison to the feedback provided to high achievers.

\section{Conclusion and Discussion}
Incorporating Generative Artificial Intelligence (GenAI), particularly Large Language Models (LLMs), into educational settings offers significant opportunities to enhance educator efficiency and enrich student learning experiences. However, adopting GenAI through conversational user interfaces (CUIs) poses significant risks and challenges \citep{kasneciChatGPTGoodOpportunities2023,choiAreLLMsUseful2023}, such as lack of AI expertise among educators \citep{oteroAILiteracyK122023}, additional responsibility for educators' to adhere to privacy guidelines and account for data leakage \citep{walkowiakGenerativeAIWorkforce2023}, and over-reliance on AI \citep{kohLearningTeachingAssessment2023,darvishiImpactAIAssistance2024}. This paper presents three main contributions to demonstrate how integrating GenAI into user-centric applications can address these challenges, offering pedagogically sound and ethically responsible approaches to using GenAI in education.

\textbf{\textit{The first contribution}} of our research is the development of a design framework that integrates prompting methodologies with pedagogical theories to create intuitive, user-centric interfaces. This framework offers a systematic method for overcoming the challenges mentioned earlier during the design phase of GenAI applications, by informing application designers on the ethical and responsible application of AI principles. To prevent potential misuse of GenAI, our framework is intended for use by designers and instructors, ensuring that those with educational expertise lead its application, thus minimizing the risk of GenAI being used incorrectly or misinterpreted \citep{oteroAILiteracyK122023}. Additionally, the framework narrows the expertise gap by offering step-by-step instructions, support for customization, and scaffolding to adapt GenAI outputs without requiring deep knowledge of prompt engineering \citep{chiuSystematicLiteratureReview2023}. Furthermore, our framework is particularly effective in supporting tailored educational interventions for individual students and enabling instructors to supervise the GenAI process. This approach allows instructors to use their time more effectively, concentrating on content accuracy. Consequently, educators have more opportunities to participate in other rewarding teaching activities, bringing a more personal element back into education. To minimise ethical and fairness concerns \citep{bondMetaSystematicReview2023}, we advise against using this framework for high-stakes decisions like grading. Instead, we advocate for its use in places such as providing immediate, detailed feedback on formative tasks that support and enhance student learning. 

\textbf{\textit{The second contribution}} of our work is the instantiation of the framework resulting in Feedback Copilot. This contribution has practical implications regarding the technical feasibility of our approach and lessons learnt from the reference implementation. We noticed that the feedback evaluation scores are skewed towards the higher end. In order to ensure a more natural distribution of evaluation scores, other prompting techniques might be used, for example, a few-shot prompting techniques with examples of low and high-scored feedback for each criterion. Recent research indicates that even though using flagship open access LLM models, e.g., GPT-4, have a high alignment with human annotations, they fall short in certain contexts \citep{zhaoEtalcross23}. This means that more research is required to understand the reliability of using GenAI as an evaluator. %
Our reference implementation integrates with existing LMS infrastructure. The current Feedback Copilot pipeline relies on pseudonymisation before using GenAI models, which changes authentic student names with random identifiers. This mechanism is aligned with the privacy standards, which is one of the major challenges with GenAI adoption \citep{yanPracticalEthicalChallenges2023}. 
We see two implications stemming from building custom-built GenAI applications for specific tasks to ensure privacy.
From the perspective of the UI design, the transition from CUI to a custom-built GenAI application could streamline the process of excluding private or sensitive data from being processed by GenAI pipelines, including situations contingent on students' decision to opt out. Our framework guides GenAI designers in providing step-by-step instructions for users to tailor GenAI outputs ethically while integrating institutional data.
In our reference implementation, we relied on the proprietary GenAI model. Enterprise GenAI currently offers limited safe access to institutional educational data. Open-source local models could help in alleviating privacy risks, however, their performance lags behind on-premise models and they require custom integrations.

\textit{\textbf{The third contribution}} of our work stems from the evaluation study, where we explored the relationship between feedback quality and feedback equitability depending on students' achievement.
Importantly, the evaluation showed that feedback generated with an advanced variation of Feedback Copilot, that included instructions provided in the prompts, had consistently better quality compared to feedback generated via a base variation, both overall and accounting for individual feedback criteria. Our results align with other research demonstrating GenAI's high-quality outcomes for instructional tasks \citep{dennyCanWeTrust2023}. 
We also found that feedback for students with lower assignment grades had lower quality. This finding has two major implications.
As indicated by \citet{ruweYourArgumentationGood2023}, students generally trust AI-provided feedback more than human-provided feedback. It means that GenAI-generated feedback could potentially match the impact of instructor-written feedback, delivering equal benefits. Building on this, the implications of GenAI-generated feedback could be profound. If students trust GenAI feedback as much as, or even more than, feedback from their instructors, it opens up new avenues for personalized learning. This could lead to a more efficient learning process where feedback is not only immediate but also tailored to the individual needs of each student.
Moreover, this finding has substantial implications for ensuring and advancing equitability of available educational resources \citep{bondMetaSystematicReview2023}. Our findings suggest that GenAI-enabled feedback could greatly benefit students, however, there is a strong need for educator oversight of generated feedback, especially for lower achieving students who often require more detailed and pedagogically valid feedback. 
In sum, this finding suggests that Feedback Copilot could alleviate instructors' workload, allowing them to focus on other instructional activities. However, it's important to ensure that the GenAI applications are designed to help effectively allocate instructors' time and effort to provide oversight that the generated feedback is accurate, constructive, and ethically sound. This will help maintain the trust of students and ensure the effectiveness of the feedback.

\textbf{\textit{Broader implications}}. The deployment of GenAI to assist educators, while holding promising potential, currently faces significant challenges. These include concerns about the reliability and accuracy of GenAI outputs, a lack of transparency and explainability in its decision-making, and broader questions regarding its suitability for educational settings.
These imperfections have far-reaching implications that span multiple dimensions, including technological, educational, ethical, psychological, and social aspects.
\textit{Technologically}, the incorporation of GenAI models necessitates robust structures to ensure data privacy, security, and ethical use, safeguarding all stakeholders in the educational process. The existing limitations highlight the need for continued research and development to enhance the sophistication and reliability of GenAI systems. This involves not only improving the accuracy of the models but also developing mechanisms that can provide users with understandable explanations for the AI's decisions and outputs.
\textit{Educationally}, the approach could elevate the quality and consistency of feedback and help teachers optimally use their time to facilitate learning. However, the current state of GenAI presents challenges in its integration into teaching and learning environments. Educators may find it difficult to trust or rely on AI-generated content or feedback due to concerns over its correctness. This skepticism can hinder the adoption of potentially transformative tools that could otherwise enhance personalized learning and instructional efficiency.
\textit{Ethically}, the deployment of imperfect GenAI tools raises questions about the fairness and consequences of their use in educational settings. Therefore, attention must be devoted to ensuring that the algorithm operates transparently and without bias, providing equitable support across diverse student populations and respecting the integrity of educational interactions.
\textit{Psychologically}, the use of GenAI in its current form can affect the attitudes and perceptions of both educators and students. Educators may feel threatened by or resistant to AI tools that seem opaque or unreliable, potentially leading to a reluctance to integrate such technologies into their teaching practices. For students, interacting with an AI system that lacks explainability could lead to confusion, frustration, and a diminished trust in the educational content being delivered.
\textit{Socially}, the integration of GenAI into educational settings has the potential to impact the dynamics of the educational community. It is essential to foster an inclusive environment that promotes collaboration and trust among all stakeholders, ensuring that GenAI tools enhance, rather than undermine, the social fabric of educational institutions. Overall, while GenAI holds significant potential, it must be implemented in a way that continues to value and foster interpersonal communication and collaborative learning among educators and students. Thus, the overarching implication is the need for a balanced, ethical, and strategically aligned incorporation of algorithms, which enhances rather than eclipses the human-centric ethos that underpins educational environments.

\ifarxiv %
    \bibliographystyle{unsrtnat}
    \bibliography{bibliography}
\else %
    \bibliographystyle{./els-cas-template/cas-model2-names}
    \bibliography{bibliography}

\begin{thebibliography}{84}
\providecommand{\natexlab}[1]{#1}
\providecommand{\url}[1]{\texttt{#1}}
\expandafter\ifx\csname urlstyle\endcsname\relax
  \providecommand{\doi}[1]{doi: #1}\else
  \providecommand{\doi}{doi: \begingroup \urlstyle{rm}\Url}\fi

\bibitem[Dickey and Bejarano(2023)]{dickeyModelIntegratingGenerative2023}
Ethan Dickey and Andres Bejarano.
\newblock A {Model} for {Integrating} {Generative} {AI} into {Course} {Content} {Development}, August 2023.
\newblock arXiv: 2308.12276 [cs] Issue: arXiv:2308.12276.

\bibitem[Denny et~al.(2023{\natexlab{a}})Denny, Khosravi, Hellas, Leinonen, and Sarsa]{dennyCanWeTrust2023}
Paul Denny, Hassan Khosravi, Arto Hellas, Juho Leinonen, and Sami Sarsa.
\newblock Can {We} {Trust} {AI}-{Generated} {Educational} {Content}? {Comparative} {Analysis} of {Human} and {AI}-{Generated} {Learning} {Resources}, July 2023{\natexlab{a}}.
\newblock arXiv: 2306.10509 [cs] Issue: arXiv:2306.10509.

\bibitem[Bulathwela et~al.(2023)Bulathwela, Muse, and Yilmaz]{bulathwelaScalableEducationalQuestion2023}
Sahan Bulathwela, Hamze Muse, and Emine Yilmaz.
\newblock Scalable {Educational} {Question} {Generation} with {Pre}-trained {Language} {Models}, May 2023.
\newblock URL \url{http://arxiv.org/abs/2305.07871}.
\newblock arXiv:2305.07871 [cs].

\bibitem[Moore et~al.(2023)Moore, Nguyen, Chen, and Stamper]{mooreAssessingQualityMultipleChoice2023}
Steven Moore, Huy~A. Nguyen, Tianying Chen, and John Stamper.
\newblock Assessing the {Quality} of {Multiple}-{Choice} {Questions} {Using} {GPT}-4 and {Rule}-{Based} {Methods}.
\newblock In Olga Viberg, Ioana Jivet, Pedro J. Muñoz-Merino, Maria Perifanou, and Tina Papathoma, editors, \emph{Responsive and {Sustainable} {Educational} {Futures}}, volume 14200, pages 229--245. Springer Nature Switzerland, Cham, 2023.
\newblock ISBN 978-3-031-42681-0 978-3-031-42682-7.
\newblock \doi{10.1007/978-3-031-42682-7_16}.
\newblock URL \url{https://link.springer.com/10.1007/978-3-031-42682-7_16}.
\newblock Series Title: Lecture Notes in Computer Science.

\bibitem[Leiker et~al.(2023)Leiker, Gyllen, Eldesouky, and Cukurova]{leikerGenerativeAILearning2023}
Daniel Leiker, Ashley~Ricker Gyllen, Ismail Eldesouky, and Mutlu Cukurova.
\newblock Generative {AI} for {Learning}: {Investigating} the {Potential} of {Learning} {Videos} with {Synthetic} {Virtual} {Instructors}.
\newblock In Ning Wang, Genaro Rebolledo-Mendez, Vania Dimitrova, Noboru Matsuda, and Olga~C. Santos, editors, \emph{Artificial {Intelligence} in {Education}. {Posters} and {Late} {Breaking} {Results}, {Workshops} and {Tutorials}, {Industry} and {Innovation} {Tracks}, {Practitioners}, {Doctoral} {Consortium} and {Blue} {Sky}}, volume 1831, pages 523--529. Springer Nature Switzerland, Cham, 2023.
\newblock ISBN 978-3-031-36335-1 978-3-031-36336-8.
\newblock \doi{10.1007/978-3-031-36336-8_81}.
\newblock URL \url{https://link.springer.com/10.1007/978-3-031-36336-8_81}.
\newblock Series Title: Communications in Computer and Information Science.

\bibitem[Pardos and Bhandari(2023)]{pardosLearningGainDifferences2023}
Zachary~A. Pardos and Shreya Bhandari.
\newblock Learning gain differences between {ChatGPT} and human tutor generated algebra hints, February 2023.
\newblock URL \url{http://arxiv.org/abs/2302.06871}.
\newblock arXiv:2302.06871 [cs].

\bibitem[Jury et~al.(2024)Jury, Lorusso, Leinonen, Denny, and Luxton-Reilly]{jury2024evaluating}
Breanna Jury, Angela Lorusso, Juho Leinonen, Paul Denny, and Andrew Luxton-Reilly.
\newblock Evaluating llm-generated worked examples in an introductory programming course.
\newblock In \emph{Proceedings of the 26th Australasian Computing Education Conference}, ACE '24, page 77–86, New York, NY, USA, 2024. Association for Computing Machinery.
\newblock ISBN 9798400716195.
\newblock \doi{10.1145/3636243.3636252}.
\newblock URL \url{https://doi.org/10.1145/3636243.3636252}.

\bibitem[Khosravi et~al.(2023)Khosravi, Denny, Moore, and Stamper]{khosraviLearnersourcingAgeAI2023}
Hassan Khosravi, Paul Denny, Steven Moore, and John Stamper.
\newblock Learnersourcing in the age of {AI}: {Student}, educator and machine partnerships for content creation.
\newblock \emph{Computers and Education: Artificial Intelligence}, 5:\penalty0 100151, 2023.
\newblock ISSN 2666920X.
\newblock \doi{10.1016/j.caeai.2023.100151}.

\bibitem[Lin et~al.()Lin, Thomas, Han, Gupta, Tan, Nguyen, and Koedinger]{linUsingLargeLanguage}
Jionghao Lin, Danielle~R Thomas, Feifei Han, Shivang Gupta, Wei Tan, Ngoc~Dang Nguyen, and Kenneth~R Koedinger.
\newblock Using {Large} {Language} {Models} to {Provide} {Explanatory} {Feedback} to {Human} {Tutors}.

\bibitem[Han et~al.(2023)Han, Yoo, Myung, Kim, Lim, Kim, Lee, Hong, Kim, Ahn, and Oh]{hanFABRICAutomatedScoring2023}
Jieun Han, Haneul Yoo, Junho Myung, Minsun Kim, Hyunseung Lim, Yoonsu Kim, Tak~Yeon Lee, Hwajung Hong, Juho Kim, So-Yeon Ahn, and Alice Oh.
\newblock {FABRIC}: {Automated} {Scoring} and {Feedback} {Generation} for {Essays}, October 2023.
\newblock URL \url{http://arxiv.org/abs/2310.05191}.
\newblock arXiv:2310.05191 [cs].

\bibitem[Choi et~al.(2023)Choi, Garrod, Atherton, Joyce-Gibbons, Mason-Sesay, and Björkegren]{choiAreLLMsUseful2023}
Jun~Ho Choi, Oliver Garrod, Paul Atherton, Andrew Joyce-Gibbons, Miriam Mason-Sesay, and Daniel Björkegren.
\newblock Are {LLMs} {Useful} in the {Poorest} {Schools}? {theTeacherAI} in {Sierra} {Leone}, October 2023.
\newblock URL \url{http://arxiv.org/abs/2310.02982}.
\newblock arXiv:2310.02982 [cs].

\bibitem[Long and Magerko(2020)]{longWhatAILiteracy2020}
Duri Long and Brian Magerko.
\newblock What is {{AI Literacy}}? {{Competencies}} and {{Design Considerations}}.
\newblock In \emph{Proceedings of the 2020 {{CHI Conference}} on {{Human Factors}} in {{Computing Systems}}}, pages 1--16, {Honolulu HI USA}, April 2020. {ACM}.
\newblock ISBN 978-1-4503-6708-0.
\newblock \doi{10.1145/3313831.3376727}.

\bibitem[White et~al.(2023)White, Fu, Hays, Sandborn, Olea, Gilbert, Elnashar, Spencer-Smith, and Schmidt]{whitePromptPatternCatalog2023}
Jules White, Quchen Fu, Sam Hays, Michael Sandborn, Carlos Olea, Henry Gilbert, Ashraf Elnashar, Jesse Spencer-Smith, and Douglas~C. Schmidt.
\newblock A {Prompt} {Pattern} {Catalog} to {Enhance} {Prompt} {Engineering} with {ChatGPT}, February 2023.
\newblock arXiv: 2302.11382 [cs] Issue: arXiv:2302.11382.

\bibitem[Cain(2023)]{cainPromptingChangeExploring2023}
William Cain.
\newblock Prompting {Change}: {Exploring} {Prompt} {Engineering} in {Large} {Language} {Model} {AI} and {Its} {Potential} to {Transform} {Education}.
\newblock \emph{TechTrends : for leaders in education \& training}, October 2023.
\newblock ISSN 8756-3894, 1559-7075.
\newblock \doi{10.1007/s11528-023-00896-0}.

\bibitem[Denny et~al.(2023{\natexlab{b}})Denny, Kumar, and Giacaman]{denny2023conversing}
Paul Denny, Viraj Kumar, and Nasser Giacaman.
\newblock Conversing with copilot: Exploring prompt engineering for solving cs1 problems using natural language.
\newblock In \emph{Proceedings of the 54th ACM Technical Symposium on Computer Science Education V. 1}, SIGCSE 2023, page 1136–1142, New York, NY, USA, 2023{\natexlab{b}}. Association for Computing Machinery.
\newblock ISBN 9781450394314.
\newblock \doi{10.1145/3545945.3569823}.
\newblock URL \url{https://doi.org/10.1145/3545945.3569823}.

\bibitem[Lim et~al.(2023)Lim, Gunasekara, Pallant, Pallant, and Pechenkina]{limGenerativeAIFuture2023}
Weng~Marc Lim, Asanka Gunasekara, Jessica~Leigh Pallant, Jason~Ian Pallant, and Ekaterina Pechenkina.
\newblock Generative {AI} and the future of education: {Ragnarök} or reformation? {A} paradoxical perspective from management educators.
\newblock \emph{The International Journal of Management Education}, 21\penalty0 (2):\penalty0 100790, July 2023.
\newblock ISSN 1472-8117.
\newblock \doi{10.1016/j.ijme.2023.100790}.
\newblock URL \url{https://www.sciencedirect.com/science/article/pii/S1472811723000289}.

\bibitem[Mollick and Mollick(2023)]{mollickAssigningAISeven2023}
Ethan Mollick and Lilach Mollick.
\newblock Assigning {AI}: {Seven} {Approaches} for {Students}, with {Prompts}, June 2023.
\newblock arXiv: 2306.10052 [cs] Issue: arXiv:2306.10052.

\bibitem[Arawjo et~al.(2023)Arawjo, Vaithilingam, Wattenberg, and Glassman]{arawjoChainForgeOpensourceVisual2023}
Ian Arawjo, Priyan Vaithilingam, Martin Wattenberg, and Elena Glassman.
\newblock {ChainForge}: {An} open-source visual programming environment for prompt engineering.
\newblock In \emph{Adjunct {Proceedings} of the 36th {Annual} {ACM} {Symposium} on {User} {Interface} {Software} and {Technology}}, pages 1--3, San Francisco CA USA, October 2023. ACM.
\newblock ISBN 9798400700965.
\newblock \doi{10.1145/3586182.3616660}.
\newblock URL \url{https://dl.acm.org/doi/10.1145/3586182.3616660}.

\bibitem[Suh et~al.(2023{\natexlab{a}})Suh, Min, Palani, and Xia]{suhSensecapeEnablingMultilevel2023}
Sangho Suh, Bryan Min, Srishti Palani, and Haijun Xia.
\newblock Sensecape: Enabling multilevel exploration and sensemaking with large language models.
\newblock In \emph{Proceedings of the 36th Annual ACM Symposium on User Interface Software and Technology}, UIST '23, pages 1--18, New York, NY, USA, 2023{\natexlab{a}}. Association for Computing Machinery.
\newblock ISBN 9798400701320.
\newblock \doi{10.1145/3586183.3606756}.
\newblock URL \url{https://doi.org/10.1145/3586183.3606756}.

\bibitem[Angert et~al.(2023)Angert, Suzara, Han, Pondoc, and Subramonyam]{angertSpellburstNodebasedInterface2023}
Tyler Angert, Miroslav~Ivan Suzara, Jenny Han, Christopher~Lawrence Pondoc, and Hariharan Subramonyam.
\newblock Spellburst: {A} {Node}-based {Interface} for {Exploratory} {Creative} {Coding} with {Natural} {Language} {Prompts}, August 2023.
\newblock arXiv: 2308.03921 [cs].

\bibitem[Kasneci et~al.(2023)Kasneci, Sessler, K{\"u}chemann, Bannert, Dementieva, Fischer, Gasser, Groh, G{\"u}nnemann, H{\"u}llermeier, Krusche, Kutyniok, Michaeli, Nerdel, Pfeffer, Poquet, Sailer, Schmidt, Seidel, Stadler, Weller, Kuhn, and Kasneci]{kasneciChatGPTGoodOpportunities2023}
Enkelejda Kasneci, Kathrin Sessler, Stefan K{\"u}chemann, Maria Bannert, Daryna Dementieva, Frank Fischer, Urs Gasser, Georg Groh, Stephan G{\"u}nnemann, Eyke H{\"u}llermeier, Stephan Krusche, Gitta Kutyniok, Tilman Michaeli, Claudia Nerdel, J{\"u}rgen Pfeffer, Oleksandra Poquet, Michael Sailer, Albrecht Schmidt, Tina Seidel, Matthias Stadler, Jochen Weller, Jochen Kuhn, and Gjergji Kasneci.
\newblock {{ChatGPT}} for good? {{On}} opportunities and challenges of large language models for education.
\newblock \emph{Learning and Individual Differences}, 103:\penalty0 102274, April 2023.
\newblock ISSN 10416080.
\newblock \doi{10.1016/j.lindif.2023.102274}.

\bibitem[Yan et~al.(2023)Yan, Sha, Zhao, Li, Martinez-Maldonado, Chen, Li, Jin, and Gašević]{yanPracticalEthicalChallenges2023}
Lixiang Yan, Lele Sha, Linxuan Zhao, Yuheng Li, Roberto Martinez-Maldonado, Guanliang Chen, Xinyu Li, Yueqiao Jin, and Dragan Gašević.
\newblock Practical and {Ethical} {Challenges} of {Large} {Language} {Models} in {Education}: {A} {Systematic} {Scoping} {Review}.
\newblock \emph{British Journal of Educational Technology}, page bjet.13370, August 2023.
\newblock ISSN 0007-1013, 1467-8535.
\newblock \doi{10.1111/bjet.13370}.
\newblock URL \url{http://arxiv.org/abs/2303.13379}.
\newblock arXiv:2303.13379 [cs].

\bibitem[Chang et~al.(2024)Chang, Wang, Wang, Wu, Yang, Zhu, Chen, Yi, Wang, Wang, Ye, Zhang, Chang, Yu, Yang, and Xie]{changSurveyEvaluationLarge2024}
Yupeng Chang, Xu~Wang, Jindong Wang, Yuan Wu, Linyi Yang, Kaijie Zhu, Hao Chen, Xiaoyuan Yi, Cunxiang Wang, Yidong Wang, Wei Ye, Yue Zhang, Yi~Chang, Philip~S. Yu, Qiang Yang, and Xing Xie.
\newblock A {{Survey}} on {{Evaluation}} of {{Large Language Models}}.
\newblock \emph{ACM Transactions on Intelligent Systems and Technology}, page 3641289, January 2024.
\newblock ISSN 2157-6904, 2157-6912.
\newblock \doi{10.1145/3641289}.

\bibitem[Bond et~al.(2023)Bond, Khosravi, De~Laat, Bergdahl, Negrea, Oxley, Pham, Chong, and Siemens]{bondMetaSystematicReview2023}
Melissa Bond, Hassan Khosravi, Maarten De~Laat, Nina Bergdahl, Violeta Negrea, Emily Oxley, Phuong Pham, Sin~Wang Chong, and George Siemens.
\newblock A {Meta} {Systematic} {Review} of {Artificial} {Intelligence} in {Higher} {Education}: {A} call for increased ethics, collaboration, and rigour.
\newblock \emph{International Journal of Educational Technology in Higher Education}, 2023.
\newblock \doi{10.13140/RG.2.2.31921.56162/1}.

\bibitem[Mazzullo et~al.(2023)Mazzullo, Bulut, Wongvorachan, and Tan]{mazzulloLearningAnalyticsEra2023}
Elisabetta Mazzullo, Okan Bulut, Tarid Wongvorachan, and Bin Tan.
\newblock Learning {Analytics} in the {Era} of {Large} {Language} {Models}.
\newblock Preprint, Social Sciences, August 2023.

\bibitem[Denny et~al.(2024)Denny, Prather, Becker, Finnie-Ansley, Hellas, Leinonen, Luxton-Reilly, Reeves, Santos, and Sarsa]{denny2024computing}
Paul Denny, James Prather, Brett~A. Becker, James Finnie-Ansley, Arto Hellas, Juho Leinonen, Andrew Luxton-Reilly, Brent~N. Reeves, Eddie~Antonio Santos, and Sami Sarsa.
\newblock Computing education in the era of generative ai.
\newblock \emph{Commun. ACM}, 67\penalty0 (2):\penalty0 56–67, jan 2024.
\newblock ISSN 0001-0782.
\newblock \doi{10.1145/3624720}.
\newblock URL \url{https://doi.org/10.1145/3624720}.

\bibitem[Afzaal et~al.(2023)Afzaal, Nouri, and Aayesha]{afzaalTransformerBasedApproachAutomatic2023}
Muhammad Afzaal, Jalal Nouri, and Aayesha Aayesha.
\newblock A {Transformer}-{Based} {Approach} for the {Automatic} {Generation} of {Concept}-{Wise} {Exercises} to {Provide} {Personalized} {Learning} {Support} to {Students}.
\newblock In Olga Viberg, Ioana Jivet, Pedro J. Muñoz-Merino, Maria Perifanou, and Tina Papathoma, editors, \emph{Responsive and {Sustainable} {Educational} {Futures}}, volume 14200, pages 16--31. Springer Nature Switzerland, Cham, 2023.
\newblock ISBN 978-3-031-42681-0 978-3-031-42682-7.
\newblock \doi{10.1007/978-3-031-42682-7_2}.
\newblock URL \url{https://link.springer.com/10.1007/978-3-031-42682-7_2}.
\newblock Series Title: Lecture Notes in Computer Science.

\bibitem[Hattie and Timperley(2007)]{hattiePowerFeedback2007}
John Hattie and Helen Timperley.
\newblock The {Power} of {Feedback}.
\newblock \emph{Review of Educational Research}, 77\penalty0 (1):\penalty0 81--112, March 2007.
\newblock ISSN 0034-6543, 1935-1046.
\newblock \doi{10.3102/003465430298487}.

\bibitem[Wisniewski et~al.(2020)Wisniewski, Zierer, and Hattie]{wisniewskiPowerFeedbackRevisited2020}
Benedikt Wisniewski, Klaus Zierer, and John Hattie.
\newblock The {{Power}} of {{Feedback Revisited}}: {{A Meta-Analysis}} of {{Educational Feedback Research}}.
\newblock \emph{Frontiers in Psychology}, 10, 2020.
\newblock ISSN 1664-1078.

\bibitem[Keuning et~al.(2019)Keuning, Jeuring, and Heeren]{keuningSystematicLiteratureReview2019}
Hieke Keuning, Johan Jeuring, and Bastiaan Heeren.
\newblock A {Systematic} {Literature} {Review} of {Automated} {Feedback} {Generation} for {Programming} {Exercises}.
\newblock \emph{ACM Transactions on Computing Education}, 19\penalty0 (1):\penalty0 1--43, March 2019.
\newblock ISSN 1946-6226.
\newblock \doi{10.1145/3231711}.

\bibitem[Cavalcanti et~al.(2021)Cavalcanti, Barbosa, Carvalho, Freitas, Tsai, Gašević, and Mello]{cavalcantiAutomaticFeedbackOnline2021}
Anderson~Pinheiro Cavalcanti, Arthur Barbosa, Ruan Carvalho, Fred Freitas, Yi-Shan Tsai, Dragan Gašević, and Rafael~Ferreira Mello.
\newblock Automatic feedback in online learning environments: {A} systematic literature review.
\newblock \emph{Computers and Education: Artificial Intelligence}, 2:\penalty0 100027, 2021.
\newblock ISSN 2666920X.
\newblock \doi{10.1016/j.caeai.2021.100027}.

\bibitem[Deeva et~al.(2021)Deeva, Bogdanova, Serral, Snoeck, and De~Weerdt]{deevaReviewAutomatedFeedback2021}
Galina Deeva, Daria Bogdanova, Estefanía Serral, Monique Snoeck, and Jochen De~Weerdt.
\newblock A review of automated feedback systems for learners: {Classification} framework, challenges and opportunities.
\newblock \emph{Computers \& Education}, 162:\penalty0 104094, March 2021.
\newblock ISSN 03601315.
\newblock \doi{10.1016/j.compedu.2020.104094}.

\bibitem[Maier and Klotz(2022)]{maierPersonalizedFeedbackDigital2022}
Uwe Maier and Christian Klotz.
\newblock Personalized feedback in digital learning environments: {Classification} framework and literature review.
\newblock \emph{Computers and Education: Artificial Intelligence}, 3:\penalty0 100080, 2022.
\newblock ISSN 2666920X.
\newblock \doi{10.1016/j.caeai.2022.100080}.
\newblock URL \url{https://linkinghub.elsevier.com/retrieve/pii/S2666920X22000352}.

\bibitem[Bernius et~al.(2022)Bernius, Krusche, and Bruegge]{berniusMachineLearningBased2022}
Jan~Philip Bernius, Stephan Krusche, and Bernd Bruegge.
\newblock Machine learning based feedback on textual student answers in large courses.
\newblock \emph{Computers and Education: Artificial Intelligence}, 3:\penalty0 100081, 2022.
\newblock ISSN 2666920X.
\newblock \doi{10.1016/j.caeai.2022.100081}.

\bibitem[Gombert et~al.(2024)Gombert, Fink, Giorgashvili, Jivet, Di~Mitri, Yau, Frey, and Drachsler]{gombertAutomatedAssessmentStudent2024}
Sebastian Gombert, Aron Fink, Tornike Giorgashvili, Ioana Jivet, Daniele Di~Mitri, Jane Yau, Andreas Frey, and Hendrik Drachsler.
\newblock From the {{Automated Assessment}} of {{Student Essay Content}} to {{Highly Informative Feedback}}: A {{Case Study}}.
\newblock \emph{International Journal of Artificial Intelligence in Education}, January 2024.
\newblock ISSN 1560-4292, 1560-4306.
\newblock \doi{10.1007/s40593-023-00387-6}.

\bibitem[Nguyen et~al.(2023)Nguyen, Stec, Hou, Di, and McLaren]{nguyenEvaluatingChatGPTDecimal2023}
Huy~A. Nguyen, Hayden Stec, Xinying Hou, Sarah Di, and Bruce~M. McLaren.
\newblock Evaluating {ChatGPT}’s {Decimal} {Skills} and {Feedback} {Generation} in a {Digital} {Learning} {Game}.
\newblock In Olga Viberg, Ioana Jivet, Pedro J. Muñoz-Merino, Maria Perifanou, and Tina Papathoma, editors, \emph{Responsive and {Sustainable} {Educational} {Futures}}, volume 14200, pages 278--293. Springer Nature Switzerland, Cham, 2023.
\newblock ISBN 978-3-031-42681-0 978-3-031-42682-7.
\newblock \doi{10.1007/978-3-031-42682-7_19}.
\newblock URL \url{https://link.springer.com/10.1007/978-3-031-42682-7_19}.
\newblock Series Title: Lecture Notes in Computer Science.

\bibitem[Jiang et~al.(2023)Jiang, Rayan, Dow, and Xia]{jiangGraphologueExploringLarge2023}
Peiling Jiang, Jude Rayan, Steven~P. Dow, and Haijun Xia.
\newblock Graphologue: {Exploring} {Large} {Language} {Model} {Responses} with {Interactive} {Diagrams}.
\newblock In \emph{Proceedings of the 36th {Annual} {ACM} {Symposium} on {User} {Interface} {Software} and {Technology}}, pages 1--20, San Francisco CA USA, October 2023. ACM.
\newblock ISBN 9798400701320.
\newblock \doi{10.1145/3586183.3606737}.
\newblock URL \url{https://dl.acm.org/doi/10.1145/3586183.3606737}.

\bibitem[Dang et~al.(2022)Dang, Mecke, Lehmann, Goller, and Buschek]{dangHowPromptOpportunities2022}
Hai Dang, Lukas Mecke, Florian Lehmann, Sven Goller, and Daniel Buschek.
\newblock How to {Prompt}? {Opportunities} and {Challenges} of {Zero}- and {Few}-{Shot} {Learning} for {Human}-{AI} {Interaction} in {Creative} {Applications} of {Generative} {Models}, September 2022.
\newblock arXiv: 2209.01390 [cs] Issue: arXiv:2209.01390.

\bibitem[Zamfirescu-Pereira et~al.(2023)Zamfirescu-Pereira, Wong, Hartmann, and Yang]{zamfirescu-pereiraWhyJohnnyCan2023}
J.D. Zamfirescu-Pereira, Richmond~Y. Wong, Bjoern Hartmann, and Qian Yang.
\newblock Why {Johnny} {Can}’t {Prompt}: {How} {Non}-{AI} {Experts} {Try} (and {Fail}) to {Design} {LLM} {Prompts}.
\newblock In \emph{Proceedings of the 2023 {CHI} {Conference} on {Human} {Factors} in {Computing} {Systems}}, pages 1--21, Hamburg Germany, April 2023. ACM.
\newblock ISBN 978-1-4503-9421-5.
\newblock \doi{10.1145/3544548.3581388}.
\newblock URL \url{https://dl.acm.org/doi/10.1145/3544548.3581388}.

\bibitem[Kim et~al.(2023{\natexlab{a}})Kim, Lee, Kim, Park, and Kim]{kimUnderstandingUsersDissatisfaction2023}
Yoonsu Kim, Jueon Lee, Seoyoung Kim, Jaehyuk Park, and Juho Kim.
\newblock Understanding {Users}' {Dissatisfaction} with {ChatGPT} {Responses}: {Types}, {Resolving} {Tactics}, and the {Effect} of {Knowledge} {Level}, November 2023{\natexlab{a}}.
\newblock URL \url{http://arxiv.org/abs/2311.07434}.
\newblock arXiv:2311.07434 [cs].

\bibitem[Kim et~al.(2023{\natexlab{b}})Kim, Lee, Shin, Kim, and Kim]{kimEvalLMInteractiveEvaluation2023}
Tae~Soo Kim, Yoonjoo Lee, Jamin Shin, Young-Ho Kim, and Juho Kim.
\newblock {EvalLM}: {Interactive} {Evaluation} of {Large} {Language} {Model} {Prompts} on {User}-{Defined} {Criteria}, September 2023{\natexlab{b}}.
\newblock URL \url{http://arxiv.org/abs/2309.13633}.
\newblock arXiv:2309.13633 [cs].

\bibitem[Sadek et~al.(2023)Sadek, Calvo, and Mougenot]{WhyCodesigningAI}
Malak Sadek, Rafael Calvo, and Céline Mougenot.
\newblock Why codesigning {AI} is different and difficult {\textbar} {ACM} {Interactions}.
\newblock \url{https://interactions.acm.org/blog/view/why-codesigning-ai-is-different-and-difficult}, 2023.
\newblock Accessed: 2023-11-06.

\bibitem[Laban et~al.(2018)Laban, Vig, Hearst, Xiong, and Wu]{labanChatExecutableVerifiable2018}
Philippe Laban, Jesse Vig, Marti~A Hearst, Caiming Xiong, and Chien-Sheng Wu.
\newblock Beyond the {Chat}: {Executable} and {Verifiable} {Text}-{Editing} with {LLMs}.
\newblock 2018.

\bibitem[Almeda et~al.(2023)Almeda, Zamfirescu-Pereira, Kim, Rathnam, and Hartmann]{PromptingDiscoveryFlexible}
Shm~Garanganao Almeda, J.~D. Zamfirescu-Pereira, Kyu~Won Kim, Pradeep~Mani Rathnam, and Bjoern Hartmann.
\newblock Prompting for {Discovery}: {Flexible} {Sense}-{Making} {forAI} {Art}-{Making} with {DreamSheets}.
\newblock 2023.

\bibitem[Huh et~al.(2023)Huh, Peng, and Pavel]{huhGenAssistMakingImage2023}
Mina Huh, Yi-Hao Peng, and Amy Pavel.
\newblock {GenAssist}: {Making} {Image} {Generation} {Accessible}, July 2023.
\newblock URL \url{http://arxiv.org/abs/2307.07589}.
\newblock arXiv:2307.07589 [cs].

\bibitem[Liu et~al.(2022)Liu, Qiao, and Chilton]{liu2022opal}
Vivian Liu, Han Qiao, and Lydia Chilton.
\newblock Opal: {Multimodal} image generation for news illustration.
\newblock In \emph{Proceedings of the 35th annual {ACM} symposium on user interface software and technology}, pages 1--17, 2022.

\bibitem[Mirowski et~al.(2023)Mirowski, Mathewson, Pittman, and Evans]{mirowskiCoWritingScreenplaysTheatre2023}
Piotr Mirowski, Kory~W. Mathewson, Jaylen Pittman, and Richard Evans.
\newblock Co-{Writing} {Screenplays} and {Theatre} {Scripts} with {Language} {Models}: {Evaluation} by {Industry} {Professionals}.
\newblock In \emph{Proceedings of the 2023 {CHI} {Conference} on {Human} {Factors} in {Computing} {Systems}}, pages 1--34, Hamburg Germany, April 2023. ACM.
\newblock ISBN 978-1-4503-9421-5.
\newblock \doi{10.1145/3544548.3581225}.
\newblock URL \url{https://dl.acm.org/doi/10.1145/3544548.3581225}.

\bibitem[Suh et~al.(2023{\natexlab{b}})Suh, Chen, Min, Li, and Xia]{suhStructuredGenerationExploration2023}
Sangho Suh, Meng Chen, Bryan Min, Toby Jia-Jun Li, and Haijun Xia.
\newblock Structured {Generation} and {Exploration} of {Design} {Space} with {Large} {Language} {Models} for {Human}-{AI} {Co}-{Creation}.
\newblock 2023{\natexlab{b}}.

\bibitem[Kim et~al.(2023{\natexlab{c}})Kim, Lee, Chang, and Kim]{kim2023cells}
Tae~Soo Kim, Yoonjoo Lee, Minsuk Chang, and Juho Kim.
\newblock Cells, generators, and lenses: {Design} framework for object-oriented interaction with large language models.
\newblock In \emph{The 36th annual {ACM} symposium on user interface software and technology ({UIST} '23), october 29-{November} 1, 2023, san francisco, {CA}, {USA}}, {UIST} '23, pages 1--18, New York, NY, USA, 2023{\natexlab{c}}. Association for Computing Machinery.
\newblock ISBN 9798400701320.
\newblock \doi{979-8-4007-0132-0/23/10}.
\newblock URL \url{https://doi.org/10.1145/3586183.3606833}.
\newblock Number of pages: 18 Place: San Francisco, CA, USA.

\bibitem[Ye et~al.(2023)Ye, Liu, Zhang, Hua, and Jia]{yeCognitiveMirageReview2023}
Hongbin Ye, Tong Liu, Aijia Zhang, Wei Hua, and Weiqiang Jia.
\newblock Cognitive {Mirage}: {A} {Review} of {Hallucinations} in {Large} {Language} {Models}, September 2023.
\newblock URL \url{http://arxiv.org/abs/2309.06794}.
\newblock arXiv:2309.06794 [cs].

\bibitem[Gero et~al.(2024)Gero, Swoopes, Gu, Kummerfeld, and Glassman]{geroSupportingSensemakingLarge2024}
Katy~Ilonka Gero, Chelse Swoopes, Ziwei Gu, Jonathan~K. Kummerfeld, and Elena~L. Glassman.
\newblock Supporting {{Sensemaking}} of {{Large Language Model Outputs}} at {{Scale}}, January 2024.

\bibitem[Chiu et~al.(2023)Chiu, Xia, Zhou, Chai, and Cheng]{chiuSystematicLiteratureReview2023}
Thomas~K.F. Chiu, Qi~Xia, Xinyan Zhou, Ching~Sing Chai, and Miaoting Cheng.
\newblock Systematic literature review on opportunities, challenges, and future research recommendations of artificial intelligence in education.
\newblock \emph{Computers and Education: Artificial Intelligence}, 4:\penalty0 100118, 2023.
\newblock ISSN 2666920X.
\newblock \doi{10.1016/j.caeai.2022.100118}.
\newblock URL \url{https://linkinghub.elsevier.com/retrieve/pii/S2666920X2200073X}.

\bibitem[Fernandez~Nieto et~al.(2022)Fernandez~Nieto, Kitto, Buckingham~Shum, and {Martinez-Maldonado}]{nieto_beyondDashboard_22}
Gloria~Milena Fernandez~Nieto, Kirsty Kitto, Simon Buckingham~Shum, and Roberto {Martinez-Maldonado}.
\newblock Beyond the {{Learning Analytics Dashboard}}: {{Alternative Ways}} to {{Communicate Student Data Insights Combining Visualisation}}, {{Narrative}} and {{Storytelling}}.
\newblock In \emph{{{LAK22}}: 12th {{International Learning Analytics}} and {{Knowledge Conference}}}, pages 219--229, {Online USA}, March 2022. {ACM}.
\newblock ISBN 978-1-4503-9573-1.
\newblock \doi{10.1145/3506860.3506895}.

\bibitem[Kirschner and {van Merri{\"e}nboer}(2013)]{kirschnerLearnersReallyKnow2013}
Paul~A. Kirschner and Jeroen~J.G. {van Merri{\"e}nboer}.
\newblock Do {{Learners Really Know Best}}? {{Urban Legends}} in {{Education}}.
\newblock \emph{Educational Psychologist}, 48\penalty0 (3):\penalty0 169--183, July 2013.
\newblock ISSN 0046-1520, 1532-6985.
\newblock \doi{10.1080/00461520.2013.804395}.

\bibitem[Luckin and Cukurova(2019)]{luckin_AIED_19}
Rosemary Luckin and Mutlu Cukurova.
\newblock Designing educational technologies in the age of {{AI}}: {{A}} learning sciences-driven approach.
\newblock \emph{British Journal of Educational Technology}, 50\penalty0 (6):\penalty0 2824--2838, November 2019.
\newblock ISSN 0007-1013, 1467-8535.
\newblock \doi{10.1111/bjet.12861}.

\bibitem[{Martinez-Maldonado} et~al.(2020){Martinez-Maldonado}, Elliott, Axisa, Power, Echeverria, and Buckingham~Shum]{maldonado_DesigningTranslucent_20}
Roberto {Martinez-Maldonado}, Doug Elliott, Carmen Axisa, Tamara Power, Vanessa Echeverria, and Simon Buckingham~Shum.
\newblock Designing translucent learning analytics with teachers: An elicitation process.
\newblock \emph{Interactive Learning Environments}, pages 1--15, January 2020.
\newblock ISSN 1049-4820, 1744-5191.
\newblock \doi{10.1080/10494820.2019.1710541}.

\bibitem[Nori et~al.(2023)Nori, Lee, Zhang, Carignan, Edgar, Fusi, King, Larson, Li, and Liu]{noriCanGeneralistFoundation2023}
Harsha Nori, Yin~Tat Lee, Sheng Zhang, Dean Carignan, Richard Edgar, Nicolo Fusi, Nicholas King, Jonathan Larson, Yuanzhi Li, and Weishung Liu.
\newblock Can {Generalist} {Foundation} {Models} {Outcompete} {Special}-{Purpose} {Tuning}? {Case} {Study} in {Medicine}.
\newblock \emph{arXiv preprint arXiv:2311.16452}, 2023.
\newblock URL \url{https://arxiv.org/abs/2311.16452}.

\bibitem[Fernando et~al.(2023)Fernando, Banarse, Michalewski, Osindero, and Rocktäschel]{fernandoPromptbreederSelfReferentialSelfImprovement2023}
Chrisantha Fernando, Dylan Banarse, Henryk Michalewski, Simon Osindero, and Tim Rocktäschel.
\newblock Promptbreeder: {Self}-{Referential} {Self}-{Improvement} {Via} {Prompt} {Evolution}, September 2023.
\newblock arXiv: 2309.16797 [cs] Issue: arXiv:2309.16797.

\bibitem[Schmidt et~al.()Schmidt, Spencer-Smith, Fu, and White]{schmidtCatalogingPromptPatterns}
Douglas~C Schmidt, Jesse Spencer-Smith, Quchen Fu, and Jules White.
\newblock Cataloging {Prompt} {Patterns} to {Enhance} the {Discipline} of {Prompt} {Engineering}.

\bibitem[Khattab et~al.(2023)Khattab, Singhvi, Maheshwari, Zhang, Santhanam, Vardhamanan, Haq, Sharma, Joshi, Moazam, Miller, Zaharia, and Potts]{khattabDSPyCompilingDeclarative2023}
Omar Khattab, Arnav Singhvi, Paridhi Maheshwari, Zhiyuan Zhang, Keshav Santhanam, Sri Vardhamanan, Saiful Haq, Ashutosh Sharma, Thomas~T. Joshi, Hanna Moazam, Heather Miller, Matei Zaharia, and Christopher Potts.
\newblock {{DSPy}}: {{Compiling Declarative Language Model Calls}} into {{Self-Improving Pipelines}}, October 2023.

\bibitem[Khosravi et~al.(2022)Khosravi, Shum, Chen, Conati, Tsai, Kay, Knight, {Martinez-Maldonado}, Sadiq, and Ga{\v s}evi{\'c}]{khosravi_XAIed_22}
Hassan Khosravi, Simon~Buckingham Shum, Guanliang Chen, Cristina Conati, Yi-Shan Tsai, Judy Kay, Simon Knight, Roberto {Martinez-Maldonado}, Shazia Sadiq, and Dragan Ga{\v s}evi{\'c}.
\newblock Explainable {{Artificial Intelligence}} in education.
\newblock \emph{Computers and Education: Artificial Intelligence}, 3:\penalty0 100074, 2022.
\newblock ISSN 2666920X.
\newblock \doi{10.1016/j.caeai.2022.100074}.

\bibitem[Longo et~al.(2024)Longo, Brcic, Cabitza, Choi, Confalonieri, Del~Ser, Guidotti, Hayashi, Herrera, Holzinger, et~al.]{longo2024explainable}
Luca Longo, Mario Brcic, Federico Cabitza, Jaesik Choi, Roberto Confalonieri, Javier Del~Ser, Riccardo Guidotti, Yoichi Hayashi, Francisco Herrera, Andreas Holzinger, et~al.
\newblock Explainable artificial intelligence (xai) 2.0: A manifesto of open challenges and interdisciplinary research directions.
\newblock \emph{Information Fusion}, page 102301, 2024.

\bibitem[Zhao et~al.(2024)Zhao, Chen, Yang, Liu, Deng, Cai, Wang, Yin, and Du]{zhao_XGenAI_24}
Haiyan Zhao, Hanjie Chen, Fan Yang, Ninghao Liu, Huiqi Deng, Hengyi Cai, Shuaiqiang Wang, Dawei Yin, and Mengnan Du.
\newblock Explainability for {{Large Language Models}}: {{A Survey}}.
\newblock \emph{ACM Transactions on Intelligent Systems and Technology}, page 3639372, January 2024.
\newblock ISSN 2157-6904, 2157-6912.
\newblock \doi{10.1145/3639372}.

\bibitem[Gao et~al.(2024)Gao, Hu, Ruan, Pu, and Wan]{gao_llmEval_24}
Mingqi Gao, Xinyu Hu, Jie Ruan, Xiao Pu, and Xiaojun Wan.
\newblock {{LLM-based NLG Evaluation}}: {{Current Status}} and {{Challenges}}, February 2024.

\bibitem[Krishna et~al.(2024)Krishna, Ramprasad, Gupta, Wallace, Lipton, and Bigham]{krishna2024genaudit}
Kundan Krishna, Sanjana Ramprasad, Prakhar Gupta, Byron~C Wallace, Zachary~C Lipton, and Jeffrey~P Bigham.
\newblock Genaudit: Fixing factual errors in language model outputs with evidence.
\newblock \emph{arXiv preprint arXiv:2402.12566}, 2024.

\bibitem[Terry et~al.(2023)Terry, Kulkarni, Wattenberg, Dixon, and Morris]{terryAIAlignmentDesign2023}
Michael Terry, Chinmay Kulkarni, Martin Wattenberg, Lucas Dixon, and Meredith~Ringel Morris.
\newblock {AI} {Alignment} in the {Design} of {Interactive} {AI}: {Specification} {Alignment}, {Process} {Alignment}, and {Evaluation} {Support}, October 2023.
\newblock URL \url{http://arxiv.org/abs/2311.00710}.
\newblock arXiv:2311.00710 [cs].

\bibitem[Subramonyam et~al.(2023)Subramonyam, Pondoc, Seifert, Agrawala, and Pea]{subramonyamBridgingGulfEnvisioning2023}
Hariharan Subramonyam, Christopher~Lawrence Pondoc, Colleen Seifert, Maneesh Agrawala, and Roy Pea.
\newblock Bridging the {Gulf} of {Envisioning}: {Cognitive} {Design} {Challenges} in {LLM} {Interfaces}, September 2023.
\newblock arXiv: 2309.14459 [cs] Issue: arXiv:2309.14459.

\bibitem[Glassman(2023)]{glassmanDesigningInterfacesHumanComputer2023}
Elena~L. Glassman.
\newblock Designing {Interfaces} for {Human}-{Computer} {Communication}: {An} {On}-{Going} {Collection} of {Considerations}, September 2023.
\newblock URL \url{http://arxiv.org/abs/2309.02257}.
\newblock arXiv:2309.02257 [cs].

\bibitem[Nicol and Macfarlane-Dick(2006)]{nicolFormativeAssessmentSelf2006}
David~J. Nicol and Debra Macfarlane-Dick.
\newblock Formative assessment and self-regulated learning: {A} model and seven principles of good feedback practice.
\newblock \emph{Studies in Higher Education}, 31\penalty0 (2):\penalty0 199--218, April 2006.
\newblock ISSN 0307-5079, 1470-174X.
\newblock \doi{10.1080/03075070600572090}.

\bibitem[Lipnevich and Panadero(2021)]{lipnevichReviewFeedbackModels2021}
Anastasiya~A. Lipnevich and Ernesto Panadero.
\newblock A {{Review}} of {{Feedback Models}} and {{Theories}}: {{Descriptions}}, {{Definitions}}, and {{Conclusions}}.
\newblock \emph{Frontiers in Education}, 6:\penalty0 720195, December 2021.
\newblock ISSN 2504-284X.
\newblock \doi{10.3389/feduc.2021.720195}.

\bibitem[Serral and Snoeck(2016)]{serralConceptualFrameworkFeedback2016}
Estefan{\'i}a Serral and Monique Snoeck.
\newblock Conceptual framework for feedback automation in sles.
\newblock In Vladimir~L. Uskov, Robert~J. Howlett, and Lakhmi~C. Jain, editors, \emph{Smart Education and E-Learning 2016}, volume~59, pages 97--107. {Springer International Publishing}, {Cham}, 2016.
\newblock ISBN 978-3-319-39689-7 978-3-319-39690-3.
\newblock \doi{10.1007/978-3-319-39690-3_9}.

\bibitem[Alfredo et~al.(2024)Alfredo, Echeverria, Jin, Yan, Swiecki, Ga{\v{s}}evi{\'c}, and Martinez-Maldonado]{alfredo2024human}
Riordan Alfredo, Vanessa Echeverria, Yueqiao Jin, Lixiang Yan, Zachari Swiecki, Dragan Ga{\v{s}}evi{\'c}, and Roberto Martinez-Maldonado.
\newblock Human-centred learning analytics and ai in education: a systematic literature review.
\newblock \emph{Computers and Education: Artificial Intelligence}, 6:\penalty0 100215, 2024.

\bibitem[Zhang(2023)]{zhangPreparingEducatorsStudents2023}
Bo~Zhang.
\newblock Preparing {Educators} and {Students} for {ChatGPT} and {AI} {Technology} in {Higher} {Education}:{Benefits}, {Limitations}, {Strategies}, and {Implications} of {ChatGPT} \&amp; {AI} {Technologies}.
\newblock 2023.
\newblock \doi{10.13140/RG.2.2.32105.98404}.
\newblock URL \url{https://rgdoi.net/10.13140/RG.2.2.32105.98404}.
\newblock Publisher: Unpublished.

\bibitem[Yang and Carless(2013)]{yangFeedbackTriangleEnhancement2013}
Min Yang and David Carless.
\newblock The feedback triangle and the enhancement of dialogic feedback processes.
\newblock \emph{Teaching in Higher Education}, 18\penalty0 (3):\penalty0 285--297, April 2013.
\newblock ISSN 1356-2517, 1470-1294.
\newblock \doi{10.1080/13562517.2012.719154}.

\bibitem[Becker et~al.(2023)Becker, Denny, {Finnie-Ansley}, {Luxton-Reilly}, Prather, and Santos]{becker_progHard_23}
Brett~A. Becker, Paul Denny, James {Finnie-Ansley}, Andrew {Luxton-Reilly}, James Prather, and Eddie~Antonio Santos.
\newblock Programming {{Is Hard}} - {{Or}} at {{Least It Used}} to {{Be}}: {{Educational Opportunities}} and {{Challenges}} of {{AI Code Generation}}.
\newblock In \emph{Proceedings of the 54th {{ACM Technical Symposium}} on {{Computer Science Education V}}. 1}, pages 500--506, {Toronto ON Canada}, March 2023. {ACM}.
\newblock ISBN 978-1-4503-9431-4.
\newblock \doi{10.1145/3545945.3569759}.

\bibitem[Field et~al.(2017)Field, Miles, and Field]{field2017discovering}
Andy Field, Jeremy Miles, and Zoe Field.
\newblock \emph{Discovering statistics using R}.
\newblock W. Ross MacDonald School Resource Services Library, 2017.

\bibitem[Tabachnick et~al.(2013)Tabachnick, Fidell, and Ullman]{tabachnick2013using}
Barbara~G Tabachnick, Linda~S Fidell, and Jodie~B Ullman.
\newblock \emph{Using multivariate statistics}, volume~6.
\newblock pearson Boston, MA, 2013.

\bibitem[Lenth(2022)]{lenthEmmeans2022}
Russell~V. Lenth.
\newblock emmeans: {Estimated} marginal means, aka least-squares means.
\newblock manual, 2022.
\newblock URL \url{https://CRAN.R-project.org/package=emmeans}.

\bibitem[{Casal-Otero} et~al.(2023){Casal-Otero}, Catala, {Fern{\'a}ndez-Morante}, Taboada, Cebreiro, and Barro]{oteroAILiteracyK122023}
Lorena {Casal-Otero}, Alejandro Catala, Carmen {Fern{\'a}ndez-Morante}, Maria Taboada, Beatriz Cebreiro, and Sen{\'e}n Barro.
\newblock {{AI}} literacy in {{K-12}}: A systematic literature review.
\newblock \emph{International Journal of STEM Education}, 10\penalty0 (1):\penalty0 29, April 2023.
\newblock ISSN 2196-7822.
\newblock \doi{10.1186/s40594-023-00418-7}.

\bibitem[Walkowiak and MacDonald(2023)]{walkowiakGenerativeAIWorkforce2023}
Emmanuelle Walkowiak and Trent MacDonald.
\newblock Generative {AI} and the {Workforce}: {What} {Are} the {Risks}?, September 2023.
\newblock URL \url{https://papers.ssrn.com/abstract=4568684}.

\bibitem[Koh and Doroudi(2023)]{kohLearningTeachingAssessment2023}
Elizabeth Koh and Shayan Doroudi.
\newblock Learning, teaching, and assessment with generative artificial intelligence: Towards a plateau of productivity.
\newblock \emph{Learning: Research and Practice}, 9\penalty0 (2):\penalty0 109--116, July 2023.
\newblock ISSN 2373-5082.
\newblock \doi{10.1080/23735082.2023.2264086}.

\bibitem[Darvishi et~al.(2024)Darvishi, Khosravi, Sadiq, Ga{\v s}evi{\'c}, and Siemens]{darvishiImpactAIAssistance2024}
Ali Darvishi, Hassan Khosravi, Shazia Sadiq, Dragan Ga{\v s}evi{\'c}, and George Siemens.
\newblock Impact of {{AI}} assistance on student agency.
\newblock \emph{Computers \& Education}, 210:\penalty0 104967, March 2024.
\newblock ISSN 03601315.
\newblock \doi{10.1016/j.compedu.2023.104967}.

\bibitem[Zhao et~al.(2023)Zhao, Xue, and Min]{zhaoEtalcross23}
Jin Zhao, Nianwen Xue, and Bonan Min.
\newblock Cross-document event coreference resolution: Instruct humans or instruct {GPT}?
\newblock In Jing Jiang, David Reitter, and Shumin Deng, editors, \emph{Proceedings of the 27th Conference on Computational Natural Language Learning (CoNLL)}, pages 561--574, Singapore, December 2023. Association for Computational Linguistics.
\newblock \doi{10.18653/v1/2023.conll-1.38}.
\newblock URL \url{https://aclanthology.org/2023.conll-1.38}.

\bibitem[Ruwe and {Mayweg-Paus}(2023)]{ruweYourArgumentationGood2023}
Theresa Ruwe and Elisabeth {Mayweg-Paus}.
\newblock ``{{Your}} argumentation is good'', says the {{AI}} vs humans {\textendash} {{The}} role of feedback providers and personalised language for feedback effectiveness.
\newblock \emph{Computers and Education: Artificial Intelligence}, 5:\penalty0 100189, 2023.
\newblock ISSN 2666920X.
\newblock \doi{10.1016/j.caeai.2023.100189}.

\end{thebibliography}
\fi
\pagebreak
\appendix

\definecolor{main-color}{rgb}{0.6627, 0.7176, 0.7764}

\lstdefinestyle{promptStyle}{
    basicstyle=\footnotesize,
    breakatwhitespace=false,         
    breaklines=true,                 
    captionpos=t,                    
    keepspaces=true,         
    frame=single,
    numbers=none,                    
    numbersep=5pt,                  
    showspaces=false,                
    showstringspaces=false,
    showtabs=false,                  
    tabsize=2
}

\lstset{style=promptStyle}

\section{Prompts used in Base and Advanced Variations of Feedback Copilot}
\label{app:prompts}

\begin{lstlisting}[caption=Base Feedback Copilot Variation: prompt pipeline consisting from step 1 and step 2 , escapechar=!, label=tab:prompts-baseline]
!\colorbox{main-color}{Step 1 -- task-wise feedback}!
You have access to: 
    i) assignment context (provided below and delineated with <assignment context> )
    ii) question context (available in the column `Task`)
    iii) sample solution (available in the column `Sample Solution`)
    iv) student's response, and (available in the column `Student Response`)
    v) student grade via rubric and overall grade for the task (available in the column `Rubric` and `Grade`)

    Here is the  assignment context:
    <assignment context> 
    We have the following schema declaration:
    \#\#\#
    Account [ username, firstName, middleName, lastName, email, dateOfBirth, registrationDate, subscriptionTier] Standard [ username ]
    Premium [ username ]
    PremiumFriend [ username, friendUsername ]
    Playlist [ username, playlistName ]
    PlaylistProduct [ username, playlistName, code ]
    Watches [ username, code, timestampStopped ]
    Product [ code, synopsis, title ]
    ProductSubtitleOptions [ code, subtitleLanguage ]
    ProductAudioOptions [ code, audioLanguage ]
    ProductTags [ code, tag ]
    Movie [ code, dateOfRelease, runtime, sequelCode]
    TVShow [ code ]
    Episode [ code, seasonNumber, episodeNumber, dateOfRelease, runtime ] Acts [ code, id ]
    ActsRoles [ code, id, roleName ]
    CastMember [ id, nationality, firstName, middleName, lastName ]

    Standard.username references Account.username
    Premium.username references Account.username
    PremiumFriend.username references Premium.username PremiumFriend.friendUsername references Premium.username
    Playlist.username references Premium.username
    PlaylistProduct.{username, playlistName} references Playlist.{username, playlistName} PlaylistProduct.code references Product.code
    Watches.username references Account.username Watches.code references Product.code ProductSubtitleOptions.code references Product.code ProductAudioOptions.code references Product.code ProductTags.code references Product.code Movie.code references Product.code Movie.sequelCode references Movie.code TVShow.code references Product.code
    Episode.code references TVShow.code Acts.code references Product.code
    Acts.id references CastMember.id ActsRoles.{code, id} references Acts.{code, id}
    \#\#\#
    </assignment context>

    User will provide to you the following information: 
    - a task, for which you will generate a feedback;
    - a sample solution for the task;
    - a student response;
    - a grade for the task;
    - a grade rubric for the task;

    There are %
    Write feedback to the student. 
    Provide detailed feedback on their response to the given question considering the provided sample solution, students' response and graded rubric. 
    Explain what has been done incorrectly. 

!\colorbox{main-color}{Step 2 -- Synthesise Feedback}!
    2 & Your task is to synthesise the task-wise feedback provided to you by the user into a single feedback for the whole assignment.
    You can elaborate on the feedback provided for each task by the user where you see fit.
    You have access to: an assignment context (provided below and delineated with <assignment context>).
    There are %

    Here is the  assignment context:
    <assignment context> 
    We have the following schema declaration:
    \#\#\#
    Account [ username, firstName, middleName, lastName, email, dateOfBirth, registrationDate, subscriptionTier] Standard [ username ]
    Premium [ username ]
    PremiumFriend [ username, friendUsername ]
    Playlist [ username, playlistName ]
    PlaylistProduct [ username, playlistName, code ]
    Watches [ username, code, timestampStopped ]
    Product [ code, synopsis, title ]
    ProductSubtitleOptions [ code, subtitleLanguage ]
    ProductAudioOptions [ code, audioLanguage ]
    ProductTags [ code, tag ]
    Movie [ code, dateOfRelease, runtime, sequelCode]
    TVShow [ code ]
    Episode [ code, seasonNumber, episodeNumber, dateOfRelease, runtime ] Acts [ code, id ]
    ActsRoles [ code, id, roleName ]
    CastMember [ id, nationality, firstName, middleName, lastName ]

    Standard.username references Account.username
    Premium.username references Account.username
    PremiumFriend.username references Premium.username PremiumFriend.friendUsername references Premium.username
    Playlist.username references Premium.username
    PlaylistProduct.{username, playlistName} references Playlist.{username, playlistName} PlaylistProduct.code references Product.code
    Watches.username references Account.username Watches.code references Product.code ProductSubtitleOptions.code references Product.code ProductAudioOptions.code references Product.code ProductTags.code references Product.code Movie.code references Product.code Movie.sequelCode references Movie.code TVShow.code references Product.code
    Episode.code references TVShow.code Acts.code references Product.code
    Acts.id references CastMember.id ActsRoles.{code, id} references Acts.{code, id}
    \#\#\#
    </assignment context> 
\end{lstlisting}

\begin{lstlisting}[caption=Advanced Feedback Copilot Variation: prompt pipeline consisting from step 1 and step 2, escapechar=!, label=tab:prompts-enhanced]
    !\colorbox{main-color}{Step 1 -- Task-Wise Feedback}!
    You have access to: 
    i) assignment context (provided below and delineated with <assignment context> )
    ii) question context (available in the column `Task`)
    iii) sample solution (available in the column `Sample Solution`)
    iv) student's response, and (available in the column `Student Response`)
    v) student grade via rubric and overall grade for the task (available in the column `Rubric` and `Grade`)

    Here is the  assignment context:
    <assignment context> 
    We have the following schema declaration:
    \#\#\#
    Account [ username, firstName, middleName, lastName, email, dateOfBirth, registrationDate, subscriptionTier] Standard [ username ]
    Premium [ username ]
    PremiumFriend [ username, friendUsername ]
    Playlist [ username, playlistName ]
    PlaylistProduct [ username, playlistName, code ]
    Watches [ username, code, timestampStopped ]
    Product [ code, synopsis, title ]
    ProductSubtitleOptions [ code, subtitleLanguage ]
    ProductAudioOptions [ code, audioLanguage ]
    ProductTags [ code, tag ]
    Movie [ code, dateOfRelease, runtime, sequelCode]
    TVShow [ code ]
    Episode [ code, seasonNumber, episodeNumber, dateOfRelease, runtime ] Acts [ code, id ]
    ActsRoles [ code, id, roleName ]
    CastMember [ id, nationality, firstName, middleName, lastName ]

    Standard.username references Account.username
    Premium.username references Account.username
    PremiumFriend.username references Premium.username PremiumFriend.friendUsername references Premium.username
    Playlist.username references Premium.username
    PlaylistProduct.{username, playlistName} references Playlist.{username, playlistName} PlaylistProduct.code references Product.code
    Watches.username references Account.username Watches.code references Product.code ProductSubtitleOptions.code references Product.code ProductAudioOptions.code references Product.code ProductTags.code references Product.code Movie.code references Product.code Movie.sequelCode references Movie.code TVShow.code references Product.code
    Episode.code references TVShow.code Acts.code references Product.code
    Acts.id references CastMember.id ActsRoles.{code, id} references Acts.{code, id}
    \#\#\#
    </assignment context>

    User will provide to you the following information: 
    - a task, for which you will generate a feedback;
    - a sample solution for the task;
    - a student response;
    - a grade for the task;
    - a grade rubric for the task;

    There are %
    Write feedback to the student. 
    Provide detailed feedback on their response to the given question considering the provided sample solution, students' response and graded rubric. 
    Explain what has been done incorrectly. & You are a friendly, helpful mentor who gives students advice and feedback about their work. 
    You need to come up with the task-wise feedback for students first. After that you will be required to provide process-based feedback. 
    Process feedback includes feedback specific to the processes underlying the tasks or relating and extending tasks. 
    Such feedback includes: recommendations about relationships among ideas; students' strategies for error detection; explicitly learning from errors; cueing the learner to different strategies and errors.

    You have access to: 
    i) assignment context (provided below and delineated with <assignment context> )
    ii) question context (available in the column `Task`)
    iii) sample solution (available in the column `Sample Solution`)
    iv) student's response, and (available in the column `Student Response`)
    v) student grade via rubric and overall grade for the task (available in the column `Rubric` and `Grade`)
    vi) feedback criteria (provided below and delineated with <feedback criteria>)

    Here is the  assignment context:
    <assignment context> 
    We have the following schema declaration:
    \#\#\#
    Account [ username, firstName, middleName, lastName, email, dateOfBirth, registrationDate, subscriptionTier] Standard [ username ]
    Premium [ username ]
    PremiumFriend [ username, friendUsername ]
    Playlist [ username, playlistName ]
    PlaylistProduct [ username, playlistName, code ]
    Watches [ username, code, timestampStopped ]
    Product [ code, synopsis, title ]
    ProductSubtitleOptions [ code, subtitleLanguage ]
    ProductAudioOptions [ code, audioLanguage ]
    ProductTags [ code, tag ]
    Movie [ code, dateOfRelease, runtime, sequelCode]
    TVShow [ code ]
    Episode [ code, seasonNumber, episodeNumber, dateOfRelease, runtime ] Acts [ code, id ]
    ActsRoles [ code, id, roleName ]
    CastMember [ id, nationality, firstName, middleName, lastName ]

    Standard.username references Account.username
    Premium.username references Account.username
    PremiumFriend.username references Premium.username PremiumFriend.friendUsername references Premium.username
    Playlist.username references Premium.username
    PlaylistProduct.{username, playlistName} references Playlist.{username, playlistName} PlaylistProduct.code references Product.code
    Watches.username references Account.username Watches.code references Product.code ProductSubtitleOptions.code references Product.code ProductAudioOptions.code references Product.code ProductTags.code references Product.code Movie.code references Product.code Movie.sequelCode references Movie.code TVShow.code references Product.code
    Episode.code references TVShow.code Acts.code references Product.code
    Acts.id references CastMember.id ActsRoles.{code, id} references Acts.{code, id}
    \#\#\#
    </assignment context>

    User will provide to you the following information: 
    - a task, for which you will generate a feedback;
    - a sample solution for the task;
    - a student response;
    - a grade for the task;
    - a grade rubric for the task;

    There are %
    Write feedback to the student. 
    Provide detailed feedback on their response to the given question considering the provided sample solution, the student has provided and the criteria. 
    Explain what has been done incorrectly. 
    The feedback you provide will be judged based on the following criteria:

    <feedback criteria>
    Constructive feedback
    Empathetic feedback 
    Feedback being detailed and actionable
    Feedback encouraging self-reflection and independence
    </feedback \vphantom{1} criteria>\\
\hline

    !\colorbox{main-color}{Step 2 -- Synthesise Feedback}!
    Your task is to synthesise the task-wise feedback provided to you by the user into a single feedback for the whole assignment.
    You can elaborate on the feedback provided for each task by the user where you see fit.
    You have access to: an assignment context (provided below and delineated with <assignment context>).
    There are %

    Here is the  assignment context:
    <assignment context> 
    We have the following schema declaration:
    \#\#\#
    Account [ username, firstName, middleName, lastName, email, dateOfBirth, registrationDate, subscriptionTier] Standard [ username ]
    Premium [ username ]
    PremiumFriend [ username, friendUsername ]
    Playlist [ username, playlistName ]
    PlaylistProduct [ username, playlistName, code ]
    Watches [ username, code, timestampStopped ]
    Product [ code, synopsis, title ]
    ProductSubtitleOptions [ code, subtitleLanguage ]
    ProductAudioOptions [ code, audioLanguage ]
    ProductTags [ code, tag ]
    Movie [ code, dateOfRelease, runtime, sequelCode]
    TVShow [ code ]
    Episode [ code, seasonNumber, episodeNumber, dateOfRelease, runtime ] Acts [ code, id ]
    ActsRoles [ code, id, roleName ]
    CastMember [ id, nationality, firstName, middleName, lastName ]

    Standard.username references Account.username
    Premium.username references Account.username
    PremiumFriend.username references Premium.username PremiumFriend.friendUsername references Premium.username
    Playlist.username references Premium.username
    PlaylistProduct.{username, playlistName} references Playlist.{username, playlistName} PlaylistProduct.code references Product.code
    Watches.username references Account.username Watches.code references Product.code ProductSubtitleOptions.code references Product.code ProductAudioOptions.code references Product.code ProductTags.code references Product.code Movie.code references Product.code Movie.sequelCode references Movie.code TVShow.code references Product.code
    Episode.code references TVShow.code Acts.code references Product.code
    Acts.id references CastMember.id ActsRoles.{code, id} references Acts.{code, id}
    \#\#\#
    </assignment context> & You are a friendly, helpful mentor who gives students advice and feedback about their work. 
    Your task is to synthesise the task-wise feedback provided to you by the user into a single feedback for the whole assignment.
    You can elaborate on the feedback provided for each task by the user where you see fit.
    There are %

    Following this, you will need to provide process-based feedback. 
    Process feedback includes feedback specific to the processes underlying the tasks or relating and extending tasks. 
    Such feedback includes: recommendations about relationships among ideas; students' strategies for error detection; explicitly learning from errors; cueing the learner to different strategies and errors.

    The feedback you provide will be judged based on the feedback criteria.

    You have access to: 
    i) assignment context (provided below and delineated with <assignment context> ) and
    ii) feedback criteria (provided below and delineated with <feedback criteria>)

    Here is the  assignment context:
    <assignment context> 
    We have the following schema declaration:
    \#\#\#
    Account [ username, firstName, middleName, lastName, email, dateOfBirth, registrationDate, subscriptionTier] Standard [ username ]
    Premium [ username ]
    PremiumFriend [ username, friendUsername ]
    Playlist [ username, playlistName ]
    PlaylistProduct [ username, playlistName, code ]
    Watches [ username, code, timestampStopped ]
    Product [ code, synopsis, title ]
    ProductSubtitleOptions [ code, subtitleLanguage ]
    ProductAudioOptions [ code, audioLanguage ]
    ProductTags [ code, tag ]
    Movie [ code, dateOfRelease, runtime, sequelCode]
    TVShow [ code ]
    Episode [ code, seasonNumber, episodeNumber, dateOfRelease, runtime ] Acts [ code, id ]
    ActsRoles [ code, id, roleName ]
    CastMember [ id, nationality, firstName, middleName, lastName ]

    Standard.username references Account.username
    Premium.username references Account.username
    PremiumFriend.username references Premium.username PremiumFriend.friendUsername references Premium.username
    Playlist.username references Premium.username
    PlaylistProduct.{username, playlistName} references Playlist.{username, playlistName} PlaylistProduct.code references Product.code
    Watches.username references Account.username Watches.code references Product.code ProductSubtitleOptions.code references Product.code ProductAudioOptions.code references Product.code ProductTags.code references Product.code Movie.code references Product.code Movie.sequelCode references Movie.code TVShow.code references Product.code
    Episode.code references TVShow.code Acts.code references Product.code
    Acts.id references CastMember.id ActsRoles.{code, id} references Acts.{code, id}
    \#\#\#
    </assignment context>

    <feedback criteria>
    Constructive feedback
    Empathetic feedback 
    Feedback being detailed and actionable
    Feedback encouraging self-reflection and independence
    </feedback criteria>\\
\end{lstlisting}

\section{Feedback Evaluation Prompt}
\label{app:evaluation-prompt}

\begin{lstlisting}[caption=Prompt template used for feedback evaluation., label=tab:eval-prompt]

Imagine that you are a feedback expert who can evaluate a given feedback based on how well it is aligned with different criteria.

1. Wait for the user input.
   User will provide you with the feedback to be provided to students.
   
2. Your role is to evaluate the feedback against the set of criteria.
    Process the criteria and their descriptions which are specified to you below:
    
    (1). Constructive Feedback:
    Feedback that highlights areas for improvement while offering specific suggestions and positive reinforcement. 
    Such feedback avoids criticism without guidance, and focuses on fostering a growth mindset.

    (2). Empathetic Feedback:
    Such feedback demonstrates understanding and sensitivity in the judgements.
    It acknowledges student's efforts and challenges, and usew language that encourages motivation rather than discouragement.

    (3). Detailed and Actionable Feedback:
    Such feedback is specific in the comments and points out particular strengths and weaknesses. 
    It offer actionable advice by breaking down larger tasks into manageable steps, facilitating the student's ability to implement changes effectively.

    (4). Encouraging Self-reflection and Independence: 
    Such feedback encourages students to think critically about their work and to become more self-directed in their learning. 
    It can include questions or suggestions that prompt students to reflect and think more deeply.
    
3. Be as strict as possible when conducting evaluations against criteria.

4. For each criterion provide a score from 0 to 10, where 0 means that the criterion is not followed at all and 10 means that the criterion is followed perfectly.

5. IMPORTANT: for each criteria give *a verbatim example* from the user-provided feedback. 
   These examples would be used to justify your evaluation.

Format output as a JSON object with the following structure:
    `{
            constructive\_feedback: ... , constructive\_feedback-justification: ...,
            empathetic\_feedback: ... , empathetic\_feedback-justification: ... ,
            detailed\_and\_actionable\_feedback: ... , detailed\_and\_actionable\_feedback-justification: ... ,
            encouraging\_self\_reflection\_and\_independence: ..., encouraging\_self\_reflection\_and\_independence-justification: ...
       }`\\
\end{lstlisting}

\section{Assignment Context: Question and Solution}
\label{app:assignment-context}

\begingroup\fontsize{7}{9}\selectfont

\begin{longtable}[t]{>{\raggedright\arraybackslash}p{3em}>{\raggedright\arraybackslash}p{30.5em}>{\raggedright\arraybackslash}p{30.5em}}
\caption{Tasks descriptions used im the assignement and a reference solution. \label{tab:task-context}}\\
\toprule
Task & Question context & Sample solution\\
\midrule
\endfirsthead
\caption[]{Tasks descriptions used im the assignement and a reference solution.  \textit{(continued)}}\\
\toprule
Task & Question context & Sample solution\\
\midrule
\endhead

\endfoot
\bottomrule
\endlastfoot
Task 1 & Find all accounts where the owner is older than 21 years old. (Note: This need to be correct at the time the query is run) & SELECT * FROM Account A WHERE YEAR(now()) - YEAR(A.dateOfBirth) >= 21\\
Task 2 & Find all playlists that do not contain a product that aired after January 1st, 2012 & SELECT playlistName FROM Playlist WHERE playlistName NOT IN( SELECT playlistName FROM PlaylistProduct WHERE PlaylistProduct.code IN( SELECT code FROM Episode WHERE dateOfRelease > '2012-1-1' UNION SELECT code FROM Movie WHERE dateOfRelease > '2012-1-1') )\\
Task 3 & Find all movies that are the third move in their franchise (i.e. at least the sequel of a sequel) & SELECT * FROM Movie A WHERE EXISTS (SELECT * FROM Movie B WHERE A.Code = B.sequelCode AND EXISTS (SELECT * FROM Movie C WHERE B.code = C.sequelCode AND NOT EXISTS( SELECT * FROM Movie D WHERE D.sequelCode = C.code)))\\
Task 4 & Find all playlists which contain a movie that the account owner has not watched & SELECT plp.playlistName FROM PlaylistProduct plp WHERE plp.code NOT IN (SELECT plp1.code FROM PlaylistProduct plp1,Watches w, Movie M WHERE plp1.username = w.username AND plp1.code=M.code)\\
Task 5 & Find which account(s) have watched at least all the products that ‘Idris Elba’ has been cast in & SELECT * FROM Account A WHERE NOT EXISTS(SELECT *     FROM CastMember C,Acts A1     WHERE C.firstName = 'Idris' AND C.lastName = 'Elba' AND A1.id = C.id AND C.id NOT IN (SELECT C.id  FROM Watches W  WHERE A1.code = W.code AND W.username = A.username))\\*
\end{longtable}
\endgroup{}

\end{document}